\documentclass[useAMS,usenatbib]{mn2e}
\usepackage[pdftex]{graphicx}
\usepackage[english]{babel}
\usepackage{amsmath}
\usepackage{amssymb}

\title[The star formation history of galaxies]{The star formation history of galaxies:
the role of galaxy mass, morphology and environment}
\author[V.Guglielmo et al.]{Valentina Guglielmo,$^{1,2}$ \thanks{E-mail:
valentina.guglielmo@oapd.inaf.it} Bianca M. Poggianti,$^{1}$ Alessia Moretti,$^{1,2}$ Jacopo Fritz,$^{3,4}$
\newauthor
Rosa Calvi,$^{5}$ Benedetta Vulcani$^{6}$, Giovanni Fasano$^{1}$, Angela Paccagnella$^{2}$\\
$^{1}$INAF-Astronomical Observatory of Padova, I-35122 Padova, Italy.\\
$^{2}$Department of Physics and Astronomy, University of Padova, I-35122 Padova, Italy.\\
$^{3}$Sterrenkundig Observatorium Vakgroep Fysica en Sterrenkunde Universiteit Gent, S9 9000 Gent, Belgium.\\
$^{4}$ Centro de Radioastronom\'\i a y Astrof\'\i sica, CRyA, UNAM, Campus Morelia, A.P. 3-72, C.P. 58089, Michoac\'an, Mexico.\\
$^{5}$ Instituto de Astrofisica de Canarias, Departamento de Astrofisica, Universidad de La Laguna, E-38200 La Laguna, Spain\\
$^{6}$Kavli Institute for the Physics and Mathematics of the Universe (WPI), The University of Tokyo Institutes for Advanced Study (UTIAS),\\ the University of Tokyo, Kashiwa, 277-8582, Japan.
}
\begin{document}

\date{Accepted: 2015 April 2. Received: 2015 March 20; in original form 2015 January 8. }

\pagerange{\pageref{firstpage}--\pageref{lastpage}} \pubyear{}

\maketitle

\label{firstpage}

\begin{abstract}
We analyze the star formation history (SFH) of galaxies as a function of present-day environment, galaxy stellar mass and morphology. The SFH is derived by means of a non-parametric spectrophotometric model applied to individual galaxies at $z \sim 0.04-0.1$ in the WIde-field Nearby Galaxy-cluster Survey (WINGS) clusters and the Padova Millennium Galaxy and Group Catalogue (PM2GC) field. The field reconstructed evolution of the star formation rate density (SFRD) follows the values observed at each redshift, except at $z>2$ where our estimate is $\sim 1.7\times$ higher than the high-z observed value. The slope of the SFRD decline with time gets progressively steeper going from low mass to high mass haloes. The decrease of the SFRD since $z=2$ is due to 1) quenching -- 50\% of the SFRD in the field and 75\% in clusters at $z>2$ originated in galaxies that are passive today -- and 2) the fact that the average SFR of today's star-forming galaxies has decreased with time. We quantify the contribution to the SFRD(z) of galaxies of today's different masses and morphologies. The current morphology correlates with the current star formation activity but is irrelevant for the past stellar history. The average SFH depends on galaxy mass, but galaxies of a given mass have different histories depending on their environment. We conclude that the variation of the SFRD(z) with environment is not driven by different distributions of galaxy masses and morphologies in clusters and field, and must be due to an accelerated formation in high mass haloes compared to low mass ones even for galaxies that will end up having the same galaxy mass today.
\end{abstract}


\section{Introduction}

In the quest to understand when galaxies formed their stars and assembled their mass, two complementary observational techniques can be employed:
direct observations of galaxies at different redshifts, and reconstruction of the previous galaxy history from fossil records at a given epoch.
The main advantage of the first method is that measuring the current star formation is less uncertain than estimating the past history, especially in galaxies in which the light of young stars outshines the older population, in particular at high redshift (Papovich et al. 2001; Zibetti et al. 2009; Conroy 2013).
On the other hand, the second method has the benefit of tracing the evolution of each individual galaxy, without having to infer histories in a statistical sense
with the problems involved in the identification of progenitors and descendants. 
Both methods heavily rely on spectrophotometric modeling, to calibrate the star formation rate (SFR) indicators and derive the star formation histories (SFHs), and are
affected by the choice of the initial mass function (IMF).

On a cosmic scale, the collection of the star formation rate density (SFRD) measurements at different cosmic times (from z=8 to 0) give us an indication on the summa of the SFH of the Universe.
(Madau et al. 1996; Lilly et al. 1996; Hopkins \& Beacom 2006; Karim et al. 2011 (radio); Burgarella et al. 2013 (FIR+UV); Sobral et al. 2013 (H-alpha); Bouwens et al. 2014 (UV); Madau \& Dickinson 2014).

It has emerged that the SFRD of the cosmos peaks at $z\sim 2$, following a rise after the Big Bang and before falling by a factor about 10 to the current value. 
This picture is now well established,
though large uncertainties still exist at high redshifts. The SFRD(z) has important implications for the reionization of the Universe, the cosmic chemical evolution, the transformation of gas into stars 
and the build-up of stellar mass. 

Ideally, however, one would want to go beyond the description of cosmic global history, and trace galaxy evolution on a galaxy-by-galaxy basis to understand the physical processes driving it.
In this respect, great progress has been made by surveys at different redshifts that have established the existence of a strong dependence of galaxy histories on galaxy stellar mass. On average, more massive galaxies have formed their stars and completed their star formation activity at higher z than less massive galaxies
(the so called downsizing effect, Cowie et al. 1996; Gavazzi et al. 2006; De Lucia et al. 2007; S\`anchez-Bl\`azquez et al. 2009).
The existence of relations between star formation rate and galaxy stellar mass (SFR-Mass) and specific star formation rate and mass (sSFR=SFR/Mass) have been established from z=0
out to $z>2$  (Brinchmann et al. 2004; Salim et al. 2007; Noeske et al. 2007; Daddi et al. 2007; Rodighiero et al. 2011; Whitaker et al. 2012; Sobral et al. 2014; Speagle et al. 2014), and many other galaxy properties have been found to be strongly correlated with galaxy mass. 
Furthermore, a number of works have pointed out that  galaxy properties are even more strongly correlated with a combination of galaxy mass and galaxy ``size'', arguing for velocity dispersion (Franx et al. 2008; Bernardi et al. 2003; Smith et al. 2009; Wake et al. 2012) or galaxy surface mass density (Brinchmann et al. 2004; Kauffmann et al.  2006) as principal drivers. The exact origin of these trends is still unknown, but evidence has accumulated for a dependence of 
galaxy stellar population ages on galaxy sizes at fixed mass (Poggianti et al. 2013; van der Wel et al. 2009; Cappellari et al. 2012), suggesting that also galaxy structure, and not just stellar mass, is relevant.
In a recent paper, Omand et al. (2014) argue that the observed correlation of the quenched fraction with $M/R^{1.5}$ is related to the dominance of the bulge component with respect to the disk, suggesting it might ultimately be linked with galaxy morphology (see also Driver et al. 2013). Even the sSFR-Mass relation might be due to the increase of the bulge mass fractions with galaxy stellar mass, as the ratio of SFR and stellar mass of the galaxy disk is virtually independent of  total stellar mass (Abramson et al. 2014).

On the other hand, galaxy stellar population properties have been known to vary strongly with galaxy environment (Spitzer \& Baade 1951; Oemler 1974; Davis \& Geller 1976; Dressler 1980). Galaxy clusters have seen an evolution in their blue galaxy fractions 
that is even stronger than in the field, and the evolution from blue star-forming to red passive takes place sooner in dense environments and massive haloes 
(Poggianti et al. 2006; Wilman et al. 2005; Cooper et al. 2006; Cucciati et al. 2006;  Iovino et al. 2010). Whether this environmental dependence is simply due to different galaxy mass distributions and/or morphological distributions with environment, or it reflects a stellar history that differs with environment at a given mass, 
is still a matter of debate (Thomas et al. 2005,2010; Baldry et al. 2006; Peng et al. 2010, 2012; Poggianti et al. 2013). On a global scale, the evolution of the SFRD in different environments at low redshift is not yet known, though the evolution of the blue galaxy fractions suggests a steeper decline in clusters than in the field (Kodama \& Bower 2001). The contribution of haloes of different masses to the SFRD(z) has been recently quantified by Popesso et al. (2014a,b), who argue that the process of structure formation, and the associated quenching processes, play an important role in the drop of the SFRD(z) since $z=1$.
Overall, several lines of evidence suggest that both galaxy mass and environment play a role, with environment being more relevant for lower mass galaxies, 
at least as far as quenching is concerned (Haines et al. 2007; Cooper et al. 2010; Pasquali et al. 2010; Peng et al. 2010, 2012; McGee et al. 2011; Sobral et al. 2011; Muzzin et al. 2012; Smith et al. 2012;  Wetzel et al. 2012; Lin et al. 2014; La Barbera et al. 2014; Vulcani et al. 2015). However, while it is well established that the relative incidence of star-forming and passive galaxies changes with environment, it is still debated whether environment matters for the whole galaxy stellar history, or it only causes it to end leading to quenching at some point.

Turning to the reconstruction of galaxy SFHs from fossil records, this reaches high levels of precision in galaxies with resolved stellar populations, such
as our Milky Way and the Local Group. Going to more distant galaxies, it has to rely on the interpretation of the galaxy integrated spectrum, and is limited by our capability
to discriminate between stars of different ages from the spectrum they emit. Spectrophotometric models capable of extracting SFHs from integrated spectra have been built by a number of groups: Heavens et al. (2000; MOPED), Cid-Fernandes et al. (2004; STARLIGHT), Ocvirk et al. (2006a,b; STECMAP), Fritz et al. (2007; now called SINOPSIS), MacArthur et al. 2009, Koleva et al. (2009; ULyss), Tojeiro et al. (2007; VESPA) and others (see sedfitting.org/SED08). They have been applied to reconstruct the SFH of galaxies in large surveys (e.g. Panter et al. 2007 and Tojeiro et al. 2009 on Sloan Digital Sky Survey (SDSS); Fritz et al. 2011 on WINGS), and to study these histories for galaxy subsets of special interest (e.g. Tojeiro et al. 2013; Vulcani et al. 2015).
Two studies in particular (Heavens et al. 2004; Panter et al. 2007) derived the cosmic SFH from SDSS spectra, and were successful in reproducing the SFRD(z) and the downsizing effect.

In this work 
we make use of a non-parametric spectrophotometric model to derive the past history of star formation in five broad bins of age from integrated spectra of galaxies in clusters and the field and, within the field, in groups and lower mass haloes. Searching for the origin of the overall decline observed in the SFRD(z) since $z=2$, we also consider present-day star-forming galaxies separately from the rest, and quantify the relative role of their decline in star formation and that of galaxies that have been quenched. 
Our goal is to shed light on the history of galaxies of different masses and morphologies, and isolate any residual environmental trend. We stress that we look for SFH trends with galaxy parameters {\sl today}, that is as a function of the mass, morphology and environment that galaxies have at low redshift, when the spectra we use to derive their past stellar history are taken.

The outline of the paper is as follows: in Sect. 2 we describe the datasets used, and in Sect. 3 the methods for assigning galaxy morphology and the spectrophotometric model used for galaxy stellar masses and SFHs. Sect. 4 presents our results: in 4.1 the SFRD of the field sample is compared with recent observational measurements at different redshifts; in 4.2 we study the SFRD in different environments; in 4.3 we analyze the SFH of star-forming galaxies both in the field and in clusters, in 4.4 the contribution of galaxies of different mass and morphological type to the total SFRD, and in 4.5 we present a global picture which considers the mean SFH of galaxies in different environments, with the same stellar mass but different morphology. Finally, we summarize our findings in Sect. 5.







The IMF adopted is a Salpeter one in the mass range 0.1-100 $M_{\odot}$ (Salpeter 1955), and the cosmological constants assumed are $\Omega_{m}$=0.3, $\Omega_{\Lambda}$=0.7, $\rm H_0=70 \, km \,s^{-1}Mpc^{-1}$.

\section{Dataset}
 \label{sec_dataset}

\subsection{PM2GC}
 \label{subsec_PM2sample}
 
The \emph{Padova Millennium Galaxy and Group Catalogue} (Calvi, Poggianti \& Vulcani, 2011) is a database built on the basis of the Millennium Galaxy Catalogue (MGC), a deep and wide B-imaging survey along an equatorial strip of $\sim$ 38 $deg^2$ obtained with the Isaac Newton Telescope (INT). The final catalogue is restricted to galaxies brighter than $\rm M_B = -18.7$ with a spectroscopic redshift in the range $\rm 0.03 \leqslant z \leqslant 0.11$, taken from the MGCz catalogue, the spectroscopic extension of the MGC, that has a 96$\%$ spectroscopic completeness at these magnitudes (Driver et al., 2005). Most of the MGCz spectra of our sample come from the SDSS (Abazajian et al. 2003, $\rm \sim 2.5 \AA$ resolution) and the remaining ones from the 2dF Galaxy Redshift Survey
(2dFGRS) (Colless et al. 2001) and the 2dF follow-up obtained by the MGC team (Driver et al. 2005), with a 2dF resolution of $\rm 9 \AA \,$ FWHM.
The fibre diameters are 3'' for the SDSS and 2.16'' for the 2dF setup, corresponding to the inner 1.3 to 6 kpc of the galaxies.
The PM2GC galaxy stellar mass completeness limit was computed as the mass of the reddest $M_B = - 18.7$ galaxy (B - V = 0.9) at our redshift upper limit (z = 0.1), and it is equal to $\rm Log M_{\star}/M_{\odot} = 10.44$.
The comoving volume of the PM2GC survey is 361424 $ \rm h^{-3} Mpc^{3}$.

The image quality and the spectroscopic completeness of the PM2GC are superior to SDSS, and these
qualities result in more robust morphological classifications and better sampling of dense regions. 
In particular, the MGC is based on INT data (2.5m telescope) obtained with a median seeing of 1.3" and at least 750s of exposure, with a pixel scale of 0.333"/pixel, while the SDSS (again, 2.5m telescope) has a median seeing of 1.5" in g (the closest band to the PM2GC), an exposure time of 54.1s and 0.396"/pixel. As for spectroscopic completeneness, 14\% of all PM2GC galaxies do not have an SDSS spectrum, and the SDSS incompleteness is particularly severe in dense regions such as groups. 
Moreover, the PM2GC 
data is very comparable in quality to our cluster sample (WINGS) and the two samples have been analyzed in a homogenous way with the same tools.

The characterization of the environment of the galaxies was conducted by means of a \emph{Friends-of-Friends} (FoF) algorithm. The methods and the presentation of the catalogues are described in Calvi et al. (2011). Briefly, a catalogue of 176 groups of galaxies with at least three members was built in the redshift range $\rm 0.04 \leqslant z \leqslant 0.1$, 
containing 43$\%$ of the total \emph{general field} population at these redshifts.
The mean redshift and velocity dispersion $\sigma$ of the groups are respectively 0.0823 and 192 $\rm km \, s^{-1}$. 88$\%$ of the selected groups are composed by less than 10 members, and 63$\%$ by less than 5 members.
Galaxies were assigned to a group if they were within 3 $\sigma$ from the group redshift and 1.5 R$_{200}$ from the group geometrical center.
We define as $R_{200}$ the radius of the sphere inside which the mean density is a factor 200 $\times$ the critical density of the Universe at that redshift. This parameter gives an approximation of the virial radius of a cluster or group and for our structures it is computed from the velocity dispersions using the formula (Finn et al., 2005):

\begin{equation}
R_{200} = 1.73 \frac{\sigma}{1000 (km/s)} \frac{1}{\sqrt{\Omega_{\Lambda}+\Omega_0 (1+z)^3}} h^{-1}  (Mpc)
\label{R200}
\end{equation}
with $\rm \sigma$ the group velocity dispersion and $z$ its mean redshift.

Galaxies that do not satisfy the group membership criteria have been placed either in the catalogue of single field galaxies, that comprises the isolated galaxies, or in the catalogue of binary field galaxies, which comprises the systems with two galaxies within 1500 km/s and 0.5 h$^{-1}$ Mpc. 
Finally, galaxies that were part of the trial groups in the FoF procedure but did not fulfill the final group membership criteria are treated separately as ``Mixed sample''.

All galaxies in the environments described above 
are collected in the "general field" sample PM2GC.

The number of galaxies in each sub-environment and in the general field sample are shown in Table~ \ref{pm2_subenv}.

\begin{table}
\begin{center}
\begin{tabular}{|c c|}
\hline
Environment & Number of galaxies\\
\hline
Groups & 1033\\
Single & 1123\\
Binary  & 486\\
Mixed Sample & 517\\
General Field & 3159\\
\hline
\end{tabular}
\end{center}
\caption{List of the number of galaxies in different environments in the PM2GC sample.}
\label{pm2_subenv}
\end{table}

In addition to the identification of PM2GC sub-environments, the masses of the dark matter haloes hosting PM2GC galaxies were estimated
by Paccagnella et al. (in preparation) 
exploiting a mock galaxy catalogue from semianalytic models (De Lucia \& Blaizot 2007) run on the Millennium Simulation (MS, Springel et al. 2005), and making use of the already mentioned FoF algorithm (Calvi et al. 2011), as described in Vulcani et al. (2014). The mass of a dark matter halo associated with a \emph{group} (where in this definition of \emph{group} also singles and binaries are included) is tightly correlated with the total stellar mass of all member galaxies (see e.g. Yang et al. 2007, 2008). Applying this method to the PM2GC magnitude limited sample, Paccagnella et al. (in prep.) derived halo masses for 1141 single galaxies, 245 binary systems and 92 groups. In this case not all PM2GC groups are considered but only 92 of the 176 in the complete catalogue, those in which the fraction of interlopers (i.e. the galaxies which are associated to a groups by the FoF algorithm due to projection effects but do not belong physically to them) is less than 30 \%.







\subsection{WINGS}

The WIde-field Nearby Galaxy-cluster Survey (WINGS) (Fasano et al., 2006) is a multi-wavelength survey of clusters at $\rm 0.04 < z < 0.07$ in the local Universe. 

The complete sample contains 76 clusters selected from three X-ray flux limited samples compiled from ROSAT All-Sky Survey data (Ebeling et al., 1996, 1998, 2000), covering a wide range in velocity dispersion, $500\,\rm km\,s^{-1} \leqslant \sigma_{cl} \leqslant 1100\,\rm km\,s^{-1}$ and X-ray luminosity, typically $\rm 0.2-5 \times 10^{44} erg\,s^{-1}$. The survey is mainly based on optical imaging in B and V bands for all the 76 clusters taken with the Wide Field Camera (WFC) mounted at the corrected f/3.9 prime focus of the INT-2.5m in La Palma and from the Wide Field Imager (WFI) at the 2.2m MPG/ESO telescope in La Silla (Varela et al., 2009). The imaging survey covers a $34' \times 34'$ field, and this area corresponds to at least 0.6R$_{200}$ for all clusters. \footnote{R$_{200}$ was computed from the cluster velocity dispersion $\sigma_{cl}$ (in km s$^{-1}$) using equation~\ref{R200} (Cava et al. 2009).}
In the following analysis all the cluster members are used regardless of clustercentric distance since the fraction of galaxies that do not satisfy the 0.6$R_{200}$ criterion is tiny compared to the entire distribution and does not affect significantly the sample.

The optical imaging was complemented by a spectroscopic survey of a subsample of about 6000 galaxies in 48 of the 76 clusters (Cava et al. 2009). The spectra were taken from August 2002 to October 2004 at the 4.2 m William Herschel Telescope (WHT) using the AF2/WYFFOS multifiber spectrograph ($\rm \sim 6\AA \,$ FWHM) and from January 2003 to March 2004 at the 3.9m Anglo Australian Telescope (AAT) using the 2dF multifiber spectrograph ($\rm \sim 9 \AA \,$ FWHM) (see Cava et al. 2009 for details). The fiber diameters were 1.6" and 2.16"  for WHT and AAT respectively, therefore the spectra cover the central 1.3 to 2.8 kpc of our galaxies depending on the cluster redshift. 
The spectroscopic selection criteria were only based on V magnitude and (B-V) colour, so to maximize the probability of observing galaxies at the cluster redshift and avoiding the introduction of biases in the sample (Cava et al., 2009). 
A galaxy is considered a member of the cluster if its spectroscopic redshift lies within $\rm \pm 3{\sigma}_{cl}$ from the cluster mean redshift.

The WINGS spectroscopic sample is affected by incompleteness. The completeness parameter, that is the ratio of the number of spectra yielding a redshift to the total number of galaxies in the parent photometric catalogue, was computed using the V-band magnitude and turned out to be essentially independent from the distance to the center of the cluster (Cava et al., 2009). In the following, SFRs and stellar mass estimates in WINGS galaxies have always been corrected for incompleteness.

From the $\sigma_{cl}$, by means of the virial theorem, the mass of the dark matter halo in which the cluster resides was calculated as follows (Poggianti et al. 2006)\footnote{This relation yields reliable mass measurements for clusters, but not for groups where the $\sigma$ is computed from a few redshifts, therefore for the groups we adopted the mass estimate method described in Sect. \ref{subsec_PM2sample}.}:

\begin{equation}
M_{halo} = 1.2 \times 10^{15} (\frac{\sigma}{1000 (km \, s^{-1})})^3 \frac{1}{\sqrt{\Omega_{\Lambda} + \Omega_0(1+z)^3}} \, h^{-1} (M_{\odot})
\label{M200}
\end{equation}

The latter equation was applied to all WINGS clusters using the velocity dispersions given in Cava et al. (2009) for 32 of the 48 clusters and for the remaining 16 clusters the most recent data from the OMEGAWINGS spectroscopic catalogue (Moretti et al. in prep.).

To compare different environments, we apply to the WINGS sample the same magnitude cut of the PM2GC.  
 Therefore, in the following, for both WINGS and PM2GC, we use only galaxies brighter than  $\rm M_B = -18.7$. In WINGS, this leaves 1249 galaxies ($\sim 2608$ when corrected
for spectroscopic incompleteness).
Equally, when considering galaxy mass bins, we will always compare WINGS and PM2GC above the same mass limit $\rm Log M_{\star}/M_{\odot} = 10.44$ (corresponding to 
$\rm M_B = -18.7$).  Only for WINGS, with no comparison in PM2GC, we will display results for an additional mass bin, down to the completeness mass limit of WINGS
which is $\rm Log M_{\star}/M_{\odot} = 10.0$.

To compute the WINGS volume, for each cluster we have considered the effective area on the sky covered by our data, derived the radius corresponding to this area, converted this radius in Mpc and computed the volume of the corresponding sphere, assuming spherical symmetry.
The total volume is the sum of the volumes of all clusters and is approximately 288 $h^{-3} Mpc^{3}$.
In order to convert this volume into the comoving value it is multiplied for a factor $(1+z)^3 = 1.17$, where z is the median redshift of the survey, z = 0.055.

\section{Methods}
 \label{sec_methods}

\subsection{Morphologies}

All galaxies in both the PM2GC and WINGS samples have been morphologically classified using MORPHOT, an automatic non parametric tool designed to obtain morphological type estimates of large galaxy samples (Fasano et al. 2007), which has been shown to be able to distinguish between ellipticals and S0 galaxies with unprecedented accuracy. It combines a set of 11 diagnostics, directly and easily computable from the galaxy image and sensitive to some particular morphological characteristic and/or feature of the galaxies. It provides two independent estimates of the morphological type based on: (i) a Maximum Likelihood technique; (ii) a Neural Network machine. The final morphological estimator combines the two techniques. The comparison with visual classifications provides an average difference in Hubble type $\Delta$T ($\leqslant$0.04) and a scatter ($\leqslant$1.7) comparable to those among visual classifications of different experienced classifiers.

The classification process has been performed using B-band images for PM2GC galaxies and V-band images for WINGS (Fasano et al. 2012), after testing that no significant systematic shift in broad morphological classification (ellipticals E, lenticulars S0 or late-types LT) exists between the V and B WINGS images (see Calvi et al. (2012) for more details). 
The morphological types we will consider are ellipticals, S0s (lenticulars) and late-types (any type later than S0s).

\subsection{SFHs and masses}
  \label{sec_spectroph_model}
  
The SFHs and stellar masses of galaxies in the PM2GC and WINGS samples are derived using a model which is an improved and extended version of the spectrophotometric code developed by Poggianti et al. (2001) to derive the SFHs from a galaxy integrated spectrum. 

The model and its application to WINGS are fully described in Fritz et al. (2007, 2011, 2014). It is based on a stellar population synthesis technique that reproduces the observed optical galaxy spectra. 

The code reproduces the main features of an observed spectrum: the equivalent widths of several lines - both in absorption and in emission - and the fluxes emitted in given bands of the continuum. 
This model assumes that an observed galactic spectrum is a combination of simple stellar population spectra, and therefore a galaxy model spectrum is computed by adding the synthetic spectra of Single Stellar Populations (SSPs) of different ages. 

The model makes use of the Padova evolutionary tracks (Bertelli et al. 1994) with AGB treatment as in Bressan et al. (1998), and
two different sets of observed stellar libraries: for ages younger than $\rm 10^9$ years Jacoby et al. (1984) was used, while for older SSPs spectra were taken from the MILES library (\rm S\`anchez-Bl\`azquez et al., 2006). Both sets were degraded in spectral resolution, in order to match that of the observed spectra. SSP spectra were then extended to the ultra-violet and infrared using theoretical libraries from Kurucz (private communication), and gas emission was included by means of the photoionization code {\small CLOUDY} (Ferland, 1996).

The initial set of SSPs was composed of 108 theoretical spectra referring to age intervals from 10$^5$ to 20 $\times$ 10$^9$ years,
that were binned into a final set of 12 SSPs used in the fitting. 

To treat dust extinction, the Galactic extinction curve ($R_v$ = 3.1, Cardelli et al. 1989) is adopted, but the value of the color excess, E(B-V) is let free to vary as a function of SSP age: dust extinction will be higher for younger stellar populations.

A single metallicity value is adopted and the model is run for three metallicities: Z = 0.05, Z = 0.02, Z = 0.004, choosing as best fit model the one with the smallest $\chi^2$. 
Fitting an observed spectrum with a single value of the metallicity is equivalent to assuming that this value belongs to the stellar population that is dominating its light. 
A check on the reliability of the mass and SFHs derived using this method has been performed analysing synthetic spectra of different SFHs with metallicity that varies as a function of stellar ages, so to simulate the chemical evolution of the galaxy, and it turns out that the way metallicity is treated does not introduce any significant bias in the recovered stellar mass or SFH (Fritz et al. 2007).

The SFH and mass estimates obtained from the fiber spectrum are scaled from the fiber magnitude to the total magnitude to recover galaxy-wide integrated
properties assuming a constant M/L. The differences in color between the fiber and the total magnitudes are however small for our cluster sample, as shown in
Fritz et al. (2011), therefore the assumption of a constant M/L ratio should not introduce large uncertainties.
It is worthwhile citing that the application of full spectral fitting techniques to integral field spectroscopy data yields much more detailed information about the SFH per pixel (ATLAS3D: Cappellari et al. 2012, CALIFA: S\'anchez et al. 2012; Cid Fernandes et al. 2013; Gonzalez-Delgado et al. 2014, SAMI: Allen et al. 2015, MaNGA: Bundy et al. 2015, CANDELS: Wuyts et al. 2012), however current Integral Field Unit (IFU) surveys are not suited for a complete census of magnitude limited samples in different 
environments.


\subsubsection{Fitting Algorithm, model outputs and uncertainties}
	\label{subsec_fitting_outputs}


During the fitting, each one of the 12 SSP spectra is multiplied by a value of SFR in that age interval.  
The fitting algorithm searches the combination of SFR values that best matches the observed spectrum, 
calculating the differences between the observed and model spectra, and evaluating them by means of a standard $\chi^2$ function.
The 12 SFR values are let free to vary completely independently from one another, without any a priori assumption on the
form of the SFH.
The observed features that are used to compare the likelihood between the model and the observed spectra are chosen from the most significant emission and absorption lines and continuum flux intervals, after the line equivalent widths are automatically measured (see Fritz et al. 2007, 2014). The observed errors on the flux are computed by taking into account the local spectral signal-to-noise ratio, while uncertainties on the equivalent widths are derived mainly from the measurements method.
An Adaptive Simulated Annealing algorithm randomly explores the parameters space, searching for the absolute minimum of the $\chi^2$ function.



The search of the combination of parameters that minimizes the differences between the observed and model spectrum is a non-linear problem and it is also underdetermined, which means that the number of constraints is lower than the number of parameters. 
The solution given with this method is non-unique, due to the limited wavelenght range under analysis, together with the age-metallicity degeneracy and the already mentioned non-linearity and underdetermination. To account for this, error bars are associated to mass, extinction and age values, computed as follows. The path performed by the minimization algorithm towards the best fit model (the minimum $\chi^2$) depends on the starting point, so, in general, starting from different initial positions can lead to different minimum points: 11 optimisations are performed, each time starting from a different point in the space parameter, obtaining 11 best fit models which are representative of the space of the solutions. Among these, the model with the median value for the mass is considered, and error bars are computed as the average difference between the values of the model with the highest and lowest total stellar mass formed in that age bin.
In this way we are confident that the expected values are contained within the error bars we calculate. The values for the stellar masses have been thoroughly compared both against other methods (Vulcani et al. 2011) and other datasets (e.g. SDSS) having objects in common with WINGS, showing an excellent agreement (Fritz et al. 2011).

The application of the spectrophotometric synthesis model allows to derive the characteristics of the stellar populations whose light constitutes the integrated spectrum: the total stellar mass, the mass of stars formed as a function of age -i.e. the SFR within each time interval in the galaxy life-, the extinction and the single "luminosity-weighted" metallicity value. It is important to keep in mind that the model outputs describe the global history of all stars that at low redshift are in the galaxy: the assembly of such stars in a single galaxy, i.e. the galaxy merger history, is totally unconstrained with this method.




All the galaxy stellar masses used in this paper are masses locked into stars, including both those that are still in the nuclear-burning phase, and remnants such as white dwarfs, neutron stars and stellar black holes.


The current SFR values are derived by fitting the flux of emission lines, whose luminosities are entirely attributed to the star formation process, neglecting all other mechanisms that can produce ionising flux. In this way, for LINERS and AGNs the SFR values can in principle be severely overestimated. The AGN identification for PM2GC galaxies was done using the latest AGN catalog from SDSS\footnote{https://www.sdss3.org/dr10/spectro/spectro\_access.php}.
The selection of AGNs in WINGS was performed with a very similar method (Marziani et al. 2013, in preparation).
We calculate that 
the AGN contribution to the total SFR of the PM2GC sample is $< 3\%$. 
Similarly, the contribution of AGNs in WINGS is only $\sim 1.6\%$. These estimates allow us to neglect the contribution of AGNs as a source of contamination in the SFRD analysed in this work (see also Sect. \ref{subsec_morphomass_plots}).

\subsubsection{Model Reliability}
 \label{subsec_model_rel}
 
The reliability of the spectrophotometric technique was tested in two ways (Fritz et al. 2007, 2011).
First, template spectra - which resemble the characteristics and the quality of the observed ones - spanning a wide range of SFHs were built,
to assess the capability of the model to  recover the input SFH. 
This test was done both on low and high S/N spectra, in order to verify whether there was a dependence of the quality of results from the spectral noise. This showed that the error bars provided by our method for the physical parameters reasonably account for the uncertainties, that are dominated by the similarity of old SSP spectra and by the limited spectral range at disposal for our analysis (Fritz et al. 2007).

The second test-phase was done on WINGS spectra in common with the SDSS project, to verify the reliability of the model in absolute terms, and
the agreement with the results on galaxy stellar masses obtained by other works was very satisfactory (Fritz et al. 2007, 2011).

There is an instrinsic degeneracy in the typical features of spectra of similar age, and this degeneracy increases for older stellar population spectra.
There is, hence, an intrinsic limit to the precision of this method in determining the age of the stellar populations that compose a spectrum. The choice of the time interval in which SFRs estimates can be considered reliable accounts for this aspect, and the initial 12 ages of the set of SSP spectra, i.e. the time intervals over which the SFR is assumed to be constant, were further binned into five intervals. These are the age intervals that are used throughout this paper. Time and corresponding redshift intervals are listed in Table \ref{zintervals_pm2gc}.

\begin{table}
\begin{center}
\begin{footnotesize}
\begin{tabular}{|c|c|c|c|c|c|c|}
\hline
\multicolumn{7}{|c|}{Model Time and Redshift intervals adopted}\\
\hline
z$_{mean}$ & z$_{lower}$ & z$_{upper}$ & $\delta_{t}$ & t$_{mean}$ & t$_{lower}$ & t$_{upper}$\\
         &           &         & Gyr & \multicolumn{3}{|c|}{Time from Big Bang (Gyr)}\\
\hline
0.06       & 0.04     & 0.09   & 0.6 & 12.7 & 12.9 & 12.3\\
0.10       & 0.09     & 0.12   & 0.4 & 11.9 & 12.3 & 11.9\\
0.40       & 0.12     & 0.67   & 4.6 & 9.6   & 11.9 & 7.3\\
1.44       & 0.67     & 2.21   & 4.4 & 5.4   & 7.3 & 2.9\\
6.49       & 2.21     & 10.71 & 2.5 & 1.4   & 2.9 & 0.4\\
\hline
\end{tabular}
\end{footnotesize}
\end{center}
\caption{Age and redshift intervals adopted. With the cosmological parameters adopted, t$_{Universe}$=13.462 Gyr. z$_{mean}$ is the mean redshift of the intervals, whose starting and ending values are given in z$_{lower}$ and z$_{upper}$ columns, respectively. $\delta_{t}$ is the corresponding time duration of the redshift bin, t$_{mean}$, t$_{lower}$ and t$_{upper}$ are the age values corresponding to z$_{mean}$, z$_{lower}$ and z$_{upper}$, respectively.}
\label{zintervals_pm2gc}
\end{table}

\begin{figure}
\begin{center}
\includegraphics[scale=0.5]{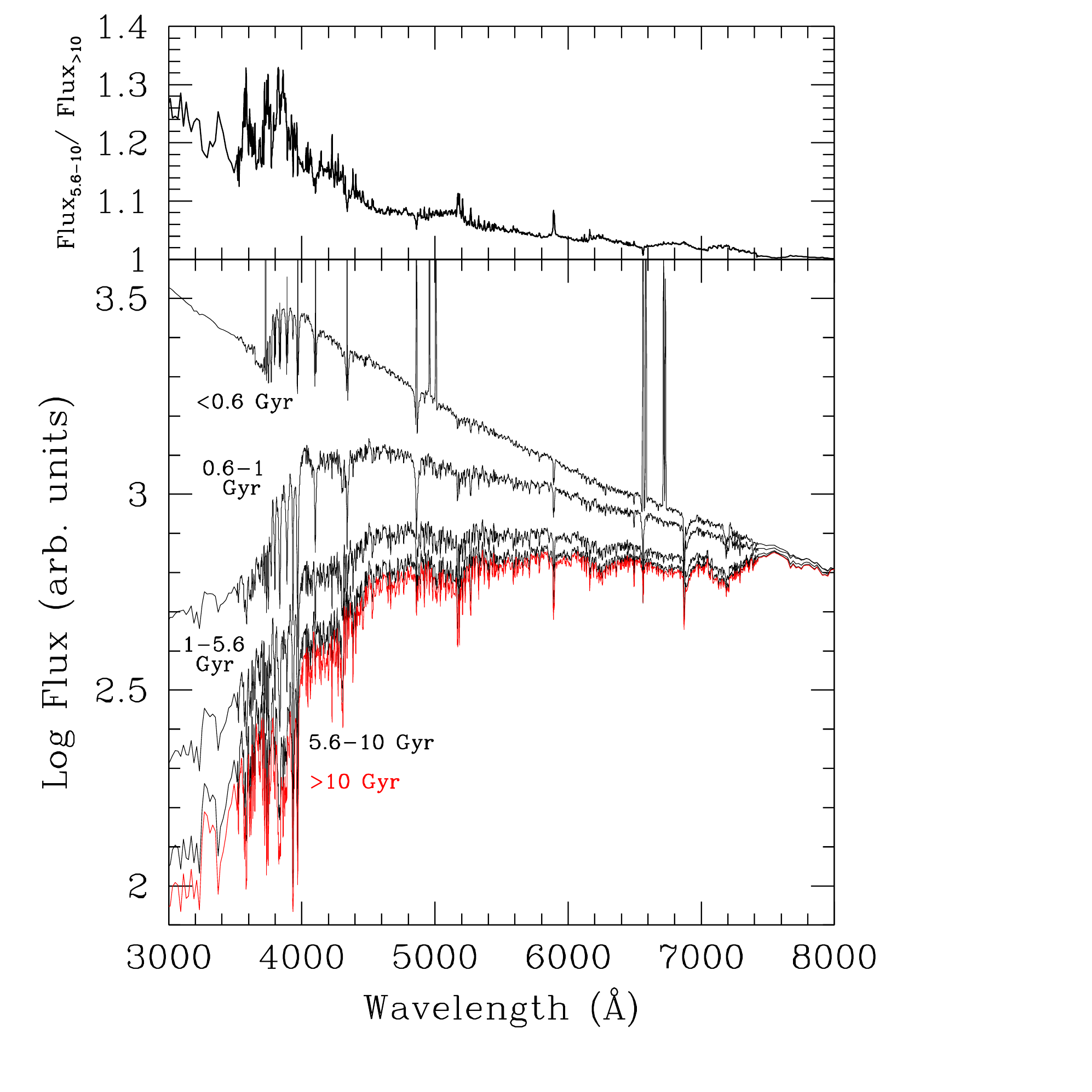}
\caption{\emph{Bottom}: Comparison between spectra of stellar populations of the five age intervals corresponding to each of the five redshift intervals in table \ref{zintervals_pm2gc}.  The age of the populations (and the redshift) is decreasing from the bottom to the top of the panel. The oldest spectrum is plotted in red. Spectra are in arbitrary units and are normalized at 8000 \AA. \emph{Top}: Ratio between the spectra of the two oldest populations.}
\label{spectra}
\end{center}
\end{figure}

To visually illustrate the reason for using a few age intervals, we plot in 
the lower panel of Figure~\ref{spectra} the spectra of stellar populations with ages reflecting the 5 age intervals adopted. The oldest spectrum, corresponding to a mean elapsed time from the Big Bang of $\sim$ 1.4 Gyr, is plotted in red in order to be distinguished from the second oldest one, which is very similar: the ratio of the fluxes of the two spectra is plotted in the upper panel of the figure, and shows 20 $\%$ level differences noticeable only in the short wavelength domain.
The plot shows that the average spectra in each time interval are significantly different one from another, and this is how their contribution to the integrated spectrum can be distinguished by the model. The only exception is the similarity between the spectra of the two oldest populations. For this reason, in the following, results at z $\geqslant$ 1 should be taken with great caution, considering the possible "spilling" between the SFR reconstruction of the two oldest populations. 

\subsubsection{Error bars on the SFR and sSFR}

When comparing model and observational SFR estimates, there are two sources of error:
that associated to the SFR estimates from the spectrophotometric model and the typical error for SFR estimates from observations.

The first type of errors, computed as described in Sect. \ref{subsec_fitting_outputs}, are considered  symmetric with respect to the central SFR value in the spectrophotometric fit.
The observational errors are taken to be equal to the typical observational error (0.225 dex), defined as the mean deviation of star formation estimates obtained using different observables (i.e UV, IR, H$\alpha$, etc) (Hao et al., 2011).
In the following, when plotting SFR estimates for WINGS and PM2GC, these two estimates are combined in quadrature.
For errors on the sSFR we calculate the propagation of errors 
assuming a typical uncertainty on the stellar mass of 0.2 dex.
The value obtained is then combined in quadrature with the observational error for the SFR, normalized with the same procedure according to the mass.

These estimates can be considered “intrinsic errors” and do not take into account eventual systematic errors arising from systematic uncertainties in the spectrophotometric modelling, for example in the single stellar population spectra due to isochrones and/or stellar libraries inaccuracy. Therefore, it is important to keep in mind that the errors shown are lower limits.

\section{Results}
 \label{sec_results}
In this section we present the methods and most significant results of the SFH analysis conducted with our spectrophotometric model on the PM2GC and WINGS. 


The reconstruction of the SFH of galaxies has been performed as follows: the twelve model SFRs are binned into the five final age intervals as described in section \ref{subsec_model_rel} computing the mean constant value of SFR for the entire length of the corresponding redshift bin. These values are then divided by the comoving volume of the survey the galaxies belong to, to obtain the SFRDs.

In all the plots five values of star formation are presented, with horizontal bars indicating the redshift interval they refer to, and vertical error bars indicating the uncertainty computed as already described.



The sub-division of galaxies according to their morphology and environments has been described in sections \ref{sec_dataset} and 	\ref{sec_methods}.
Below, we will consider the following galaxy stellar mass bins: 
\begin{itemize}
\item $\rm 10 \leqslant log M_{star} (M_{\odot}) < 10.44$ - this bin is used only for WINGS galaxies, whose mass completeness limit is lower than in the PM2GC
\item M1: $\rm 10.44 \leqslant log M_{star} (M_{\odot}) < 10.7$
\item M2: $\rm 10.7 \leqslant log M_{star} (M_{\odot}) < 11.2$
\item M3: $\rm log M_{star} (M_{\odot}) \geqslant 11.2$. The most massive galaxy in the PM2GC has $\rm log M_{star} (M_{\odot}) = 12.6$, and in WINGS $\rm log M_{star} (M_{\odot}) = 12.5$.
\end{itemize}

In the following, it is important to keep in mind that there is a degeneracy between
the results in the two highest redshift bins. However, the total stellar mass formed 
at these high redshifts is well constrained, since it is strictly linked to the 
observed spectrum and the stellar mass formed at lower
redshifts.

\subsection{The cosmic SFH}

\begin{figure}
\includegraphics[scale=0.425]{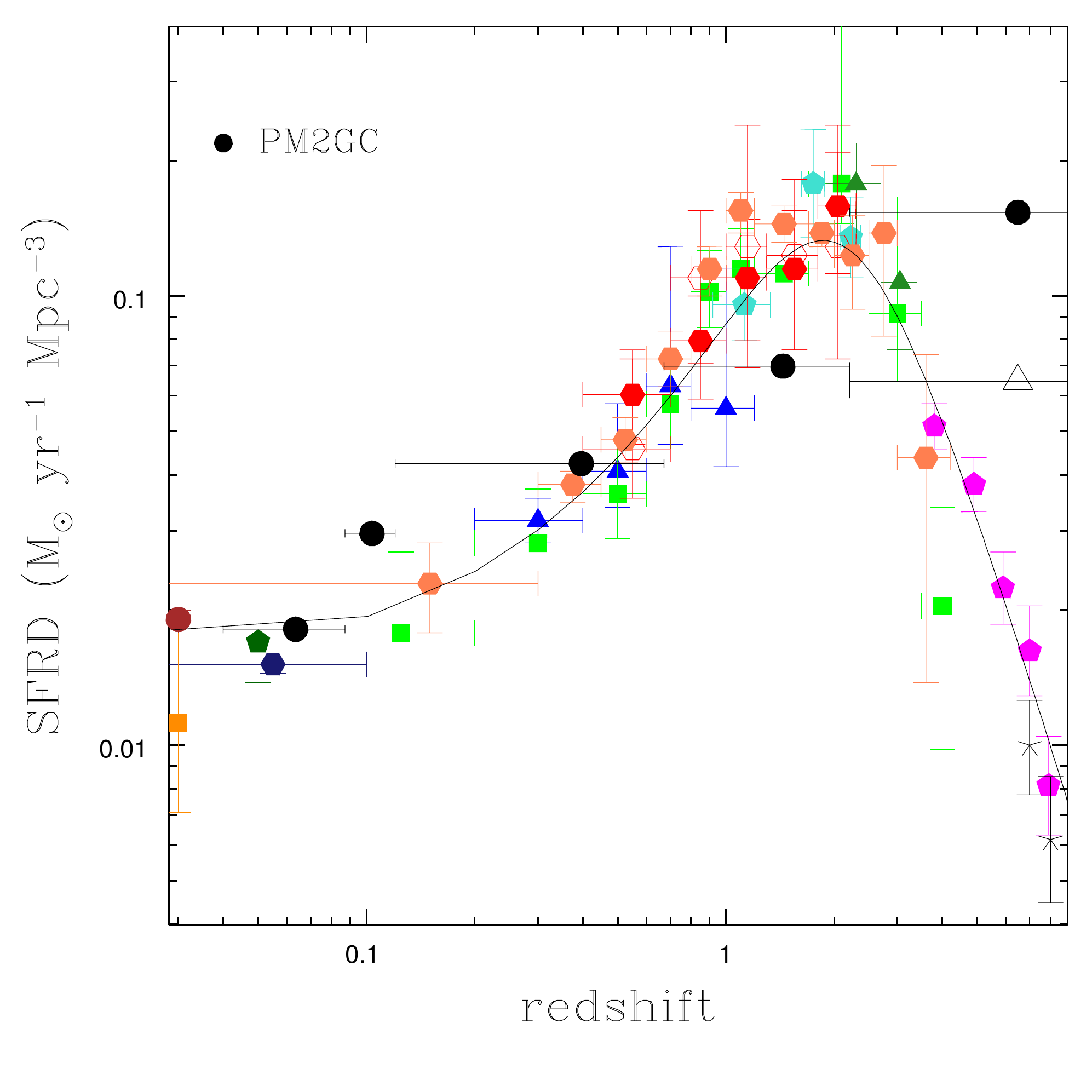}
\caption{Comparison between the PM2GC cosmic SFH and observational data from the literature (Table~1 in MD14). The black circles refer to the PM2GC field dataset. Error bars in ordinate are smaller than the symbols, while the horizontal error bars show the redshift intervals each circle is referring to. The solid curve is the best-fit SFRD shown in equation \ref{eqMadau}, as calculated by MD14. The black empty  triangle is the integral of the MD14 curve between 10 Gyr and 13 Gyr, corresponding to the last redshift bin in PM2GC. The data points refer to FUV+UV  and mid- and far-IR rest-frame measurements and are taken from Table~1 in MD14. Wyder et al. (2005), midnight blue hexagon. Schiminovich et al. (2005), blue triangles. Robotham $\&$ Driver (2011), dark green pentagon. Cucciati et al. (2012), green squares. Dahlen et al. (2007), turquoise pentagons. Reddy $\&$ Steidel (2009), forest green triangles. Bouwens et al. (2012a),(2012b), magenta pentagons. Schenker et al. (2013), black crosses. Sanders et al. (2003), brown circle. Takeuchi et al. (2003), dark orange square. Magnelli et al. (2011), red open hexagons. Magnelli et al. (2013), red filled hexagons. Gruppioni et al. (2013), coral hexagons.}
\label{Pm2gc_MD14}
\end{figure}

In Fig. \ref{Pm2gc_MD14} we compare the SFRD of the PM2GC general field sample and the cosmic SFH derived from the most recent data at all redshifts  (Madau $\&$ Dickinson 2014, MD14).
These latter data are taken from galaxy surveys that provide SFR measurements from rest-frame far-UV (1500 \AA) and mid- and far-infrared, and span the redshift range $z = 0-8$. All the surveys considered provide best fit LF parameters, therefore SFRD values can be obtained integrating the luminosity functions down to the same limiting luminosity in units of the characteristic luminosity L$^*$, $L_{min} = 0.03 L^*$. A Salpeter$_{0.1-100}$ IMF was assumed in MD14.
Together with the data we also plot the best-fitting function given by MD14, expressed by the analytical form:

\begin{equation}
SFRD(z) = 0.015 \frac{(1+z)^{2.7}}{1+[(1+z)/2.9]^{5.6}} M_{\odot} yr^{-1} Mpc^{-3}.
\label{eqMadau}
\end{equation}

The PM2GC values are shown as black circles. We note that the Madau and PM2GC values 
refer to galaxy samples selected with different criteria: the $L_{min} = 0.03 L^*$ limit at each redshift in MD14, as opposed to $M_B<-18.7$ in the PM2GC at low-z.
The PM2GC SFRD trend follows quite well the SFRD estimates at different redshifts, suggesting that the histories traced by galaxies selected according to the PM2GC criterion account quite well for the cosmic evolution derived adopting the MD14 selection.\footnote{To assess the effect of the different selection criteria on the total SFRD estimate at low redshift,
we compare the integral of the PM2GC SFR distribution function for the $M_B=-18.7$ limited
sample with that of the SFR function measured by Bothwell et al. (2011) for galaxies at z=0.005-0.1 
integrated down to 0.03$L^{\star}$. We find that the $M_B=-18.7$ cut yields a total SFR value that
is 9\% higher than the 0.03$L^{\star}$ cut, thus we conclude that the different criteria
can lead to a $\sim 10$\% difference.} The most noticeable discrepancy is in the highest redshift bin ($z>2$), where the PM2GC value is a factor $\sim 1.66$ higher than the mean SFRD obtained by integrating the MD14 best-fit function at the same epoch.
This behaviour can have several reasons:
a) the uncertainty in the two highest redshift bins of the SFRD computed by our spectrophotometric model already discussed in section \ref{sec_spectroph_model}; b) an underestimation of the observed SFRD due to incompleteness of high redshift data from current surveys; c) the differences in the MD14 vs. PM2GC selection criteria mentioned above.


\subsection{The SFH in different environments}
 \label{subsec_environ_plots}

\begin{figure}
\includegraphics[scale=0.425]{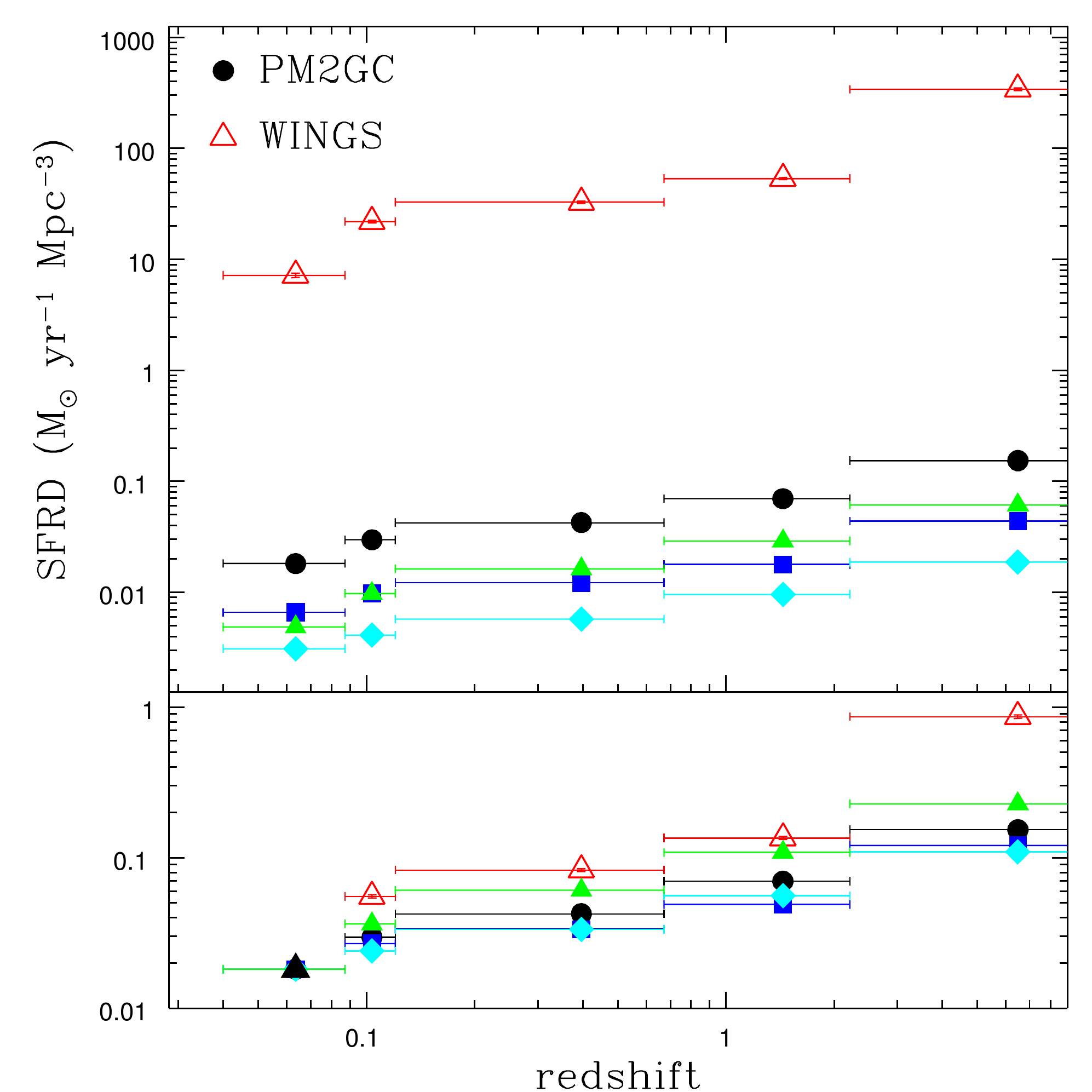}
\caption{Comparison between the field (PM2GC, black circles) and clusters (WINGS, red empty triangles) SFRD. The field sample has also been divided into groups (green triangles), binary (cyan diamonds) and single (blue squares) galaxies. Horizontal bars show the extension of time the circles are referring to. In the top panel SFRD is given in $\rm M_{\odot} \, yr^{-1} \, Mpc^{-3}$, in the bottom panel all samples are normalized to the PM2GC low-z value, indicated by the black solid triangle.}
\label{pm2wings_subsets}
\end{figure}

In figure~\ref{pm2wings_subsets} we compare the PM2GC field SFRD (black circles) with that of the WINGS cluster sample (red empty triangles). The PM2GC sample has been also divided into single galaxies (blue squares), binaries (cyan diamonds) and groups (green full triangles), according to the criteria described in Sect. \ref{sec_dataset}.

The SFRD is systematically higher in clusters than in the field, of a factor $>100$ at any redshift. This simply reflects the difference in density (number of galaxies per unit volume) between the two environments, being clusters much denser environments than the field. Single galaxies contribute to the total field SFRD by a factor 1.4 higher than groups in the lowest redshift bin, while at $z>0.1$ the relation is inverted in favour of groups by about the same factor. 

To better compare the slope of the SFRD in the different environments, in the lower panel all the values have been normalized so to coincide with the SFRD of PM2GC at z=0. It is evident that in all environments the star formation process was more active in the past than at the present age, which makes the SFRDs decrease with decreasing redshift. However, the slope is much steeper for cluster than for field galaxies. 
Clusters have formed the majority of their stars at high z: 2/3 of all stars ever formed in clusters were born at $z\geqslant 2$, while more than half of all stars in field galaxies formed at $z<2$. The decreasing factor defined as the ratio of  SFRD in the highest and the lowest redshift bin is roughly 40 for WINGS, while is $\sim$ 7 for PM2GC.
Considering the field finer environments, in groups the decreasing factor is
10.5, while it is
5 and 5.5 for binary and single galaxies, respectively. 

Calvi et al. (2013) found that the galaxy stellar mass function is similar in the field and in clusters at these magnitudes/masses,
therefore the different slope of the decline of the SFRD with redshift in the clusters and in the field is not due to the presence in clusters of more massive galaxies whose star formation occurred at earlier epochs compared to lower mass galaxies.
We will return to this point in more detail in the next sections.

\begin{figure}
\includegraphics[scale=0.425]{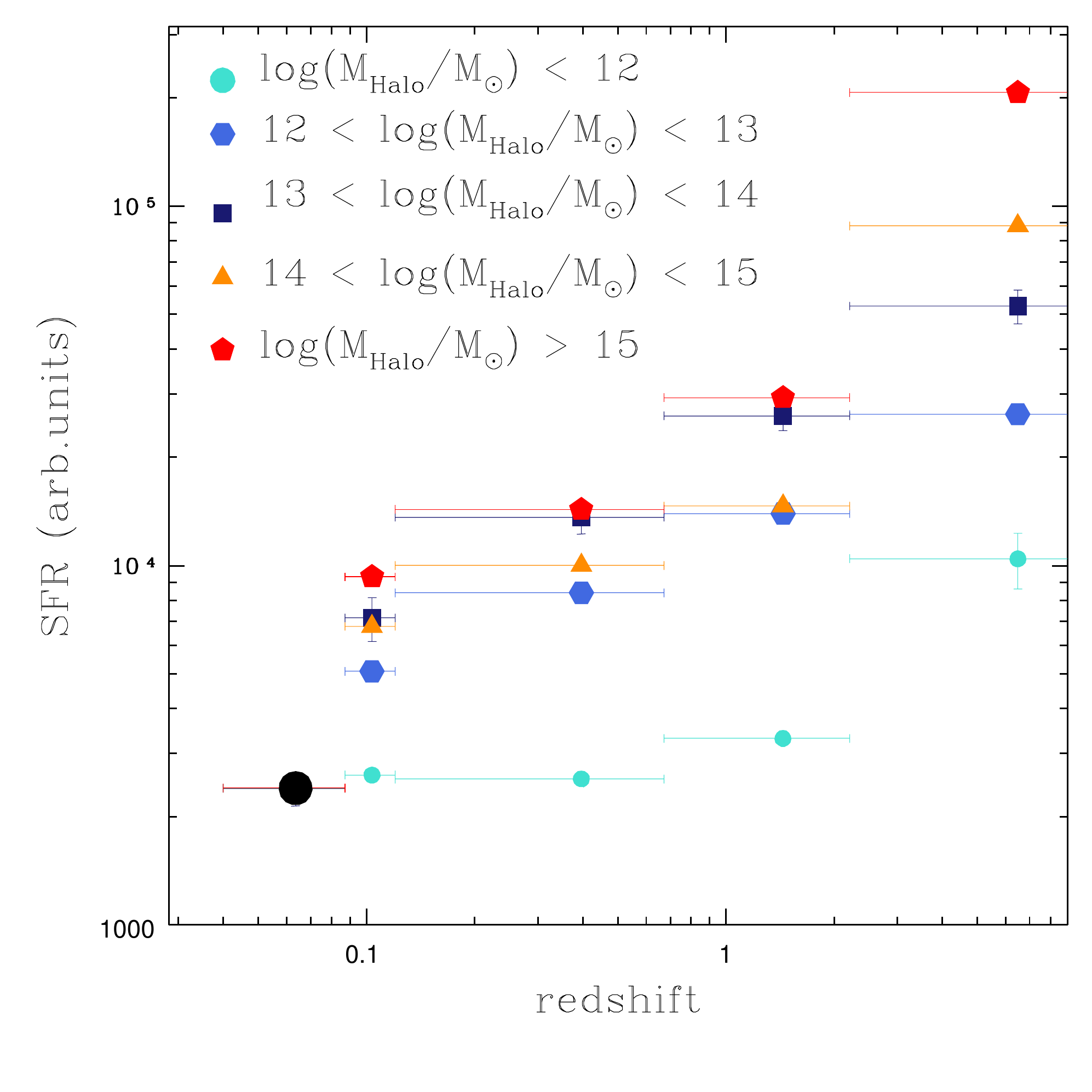}
\caption{The SFH of galaxies divided according to the mass of their host halo. Galaxies are from the PM2GC sample until halo masses of $log(M_{Halo})/M_{\odot}$=14 and from the WINGS cluster sample for more massive halos. The SFRs are normalized so to coincide in the lowest redshift bin, as indicated by the large black filled circle. The halo mass ranges considered are shown in the legend.}
\label{pm2wings_halomasses}
\end{figure}

In figure \ref{pm2wings_halomasses} the SFH of PM2GC and WINGS galaxies is plotted according to the mass of their parent dark matter halo. 
In the PM2GC, only galaxies in systems with halo masses $\rm < 10^{14} M_{\odot}$ are considered, while in WINGS only galaxies in more massive systems are taken into account. 
The number of galaxies in each halo mass interval is listed in Table~\ref{table_HM}.
The SFRs on the y-axis are normalized so to be equal to that of galaxies in the lowest mass haloes in the lowest redshift bin. As a consequence, only the redshift dependence of the SFHs of galaxies hosted in different halos is compared, while absolute values of SFR are not. 

\begin{table}
\begin{center}
\begin{tabular}{|c|c|c|}
\hline
Data sample & Halo Mass & Number of galaxies\\
\hline
PM2GC & $\rm M_{halo} < 10^{12} M_{\odot}$  & 1137 \\
PM2GC & $\rm 10^{12} M_{\odot} < M_{halo} < 10^{13} M_{\odot}$ & 708\\
PM2GC & $\rm 10^{13} M_{\odot} < M_{halo} < 10^{14} M_{\odot}$ & 261\\
WINGS & $\rm 10^{14} M_{\odot} < M_{halo} < 10^{15} M_{\odot}$ & 771\\
WINGS & $\rm M_{halo} > 10^{15} M_{\odot}$ & 478\\
\hline
\end{tabular}
\end{center}
\caption{List of the number of galaxies with different halo mass estimates both in the PM2GC and WINGS samples.}
\label{table_HM}
\end{table}

Globally, the decline in SFH gets progressively steeper going from lower to higher mass haloes. The exact shape of such decline seems to vary with the halo mass. In fact, the SFR ranking order at the highest redshift respects the halo mass ranking, while at $z \sim 0.1-2$ the ranking of 
$10^{13}-10^{14} M_{\odot}$ groups 
and $10^{14}-10^{15} M_{\odot}$ clusters is not respected.
Galaxies in haloes of mass $< \rm 10^{12} M_{\odot}$ show a very flat and well separated SFH compared to all other masses. Yet, they still display a SFR
at $z>2$ significantly higher than at any other redshift, a feature common to haloes of all masses.

To conclude, the slope of the decline of the SFRD strongly changes with environment. This is clearly visible with both of our definitions of environment. In Sect. 4.5 we will analyze in detail the origin of this effect.


\subsection{The SFH of star-forming galaxies}

The SFH throughout the cosmic time in a given environment includes a large number of galaxies and at each epoch is the result of star formation processes taking place in galaxies that are still actively forming stars. 
The decline of the SFRD from the past to the present age is in principle the cumulative result of declining star formation in galaxies that are still star-forming today (i.e. at the redshift they are observed in the PM2GC or WINGS) together with the increase in the number of galaxies that at some point have stopped forming stars, i.e. have been quenched.
The study of the SFH of today's star-forming galaxies aims to disentangle these two effects.

In the following, we consider as currently star-forming those galaxies whose sSFR at the time they are observed (i.e. z = 0.03-0.11) is above a fixed threshold. For computing the sSFR, the current SFR is taken to be the average during the last 20 Myr as obtained from the model. 

\begin{figure}
\includegraphics[scale=0.55]{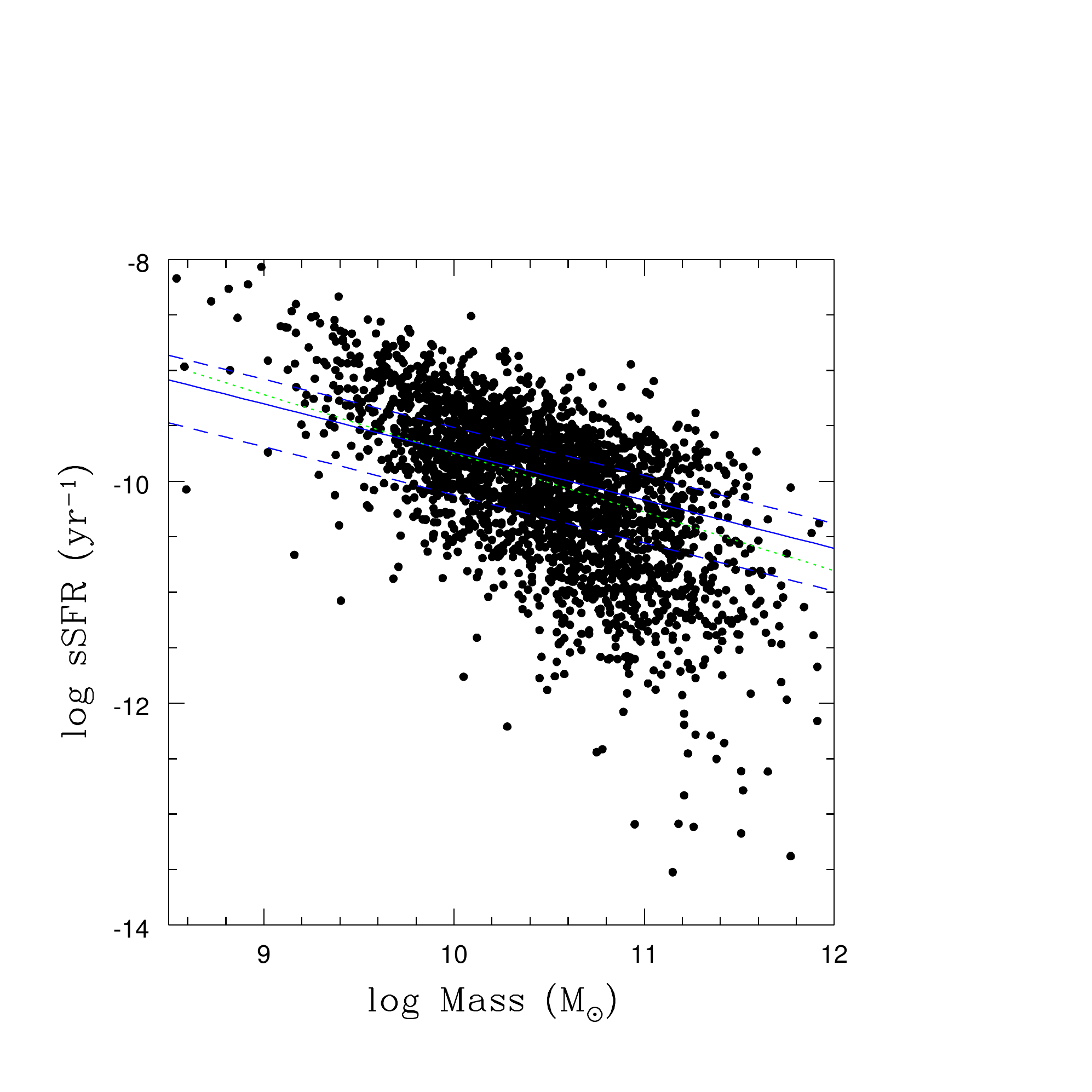}
\caption{The sSFR-Mass relation. Black dots refer to the PM2GC galaxy sample. The main sequence of star forming galaxies from Salim et al. (2007) is plotted with the green dotted line and from Lara-Lopez et al (2013) with the blue solid line. The two blue dashed lines are located at one sigma with respect to the blue solid one.}
\label{ssfrmass_rel}
\end{figure}

In figure \ref{ssfrmass_rel} we report the sSFR-Mass relation from low redshift measurements of SFR and galaxy stellar masses. The black dots in the figure are the PM2GC field galaxy sample values (z = 0.03-0.11), the green dotted line is the fit from the star forming sequence from Salim et al. (2007) (z $\simeq$ 0.1), the blue solid line is the same quantity as given in Lara-Lopez et al. (2013) (z up to $\simeq$ 0.36) and the blue dashed lines are located at one sigma with respect to the blue solid one. 
The threshold separating star-forming from passive galaxies is chosen on the basis of this sSFR-mass relation and is taken to be equal to sSFR $= 10^{-12} yr^{-1}$ (see Fig.~\ref{ssfrmass_rel}). 
This criterion selects 2094 star-forming galaxies in the field and 612 in clusters.

\begin{figure*}
\includegraphics[scale=0.425]{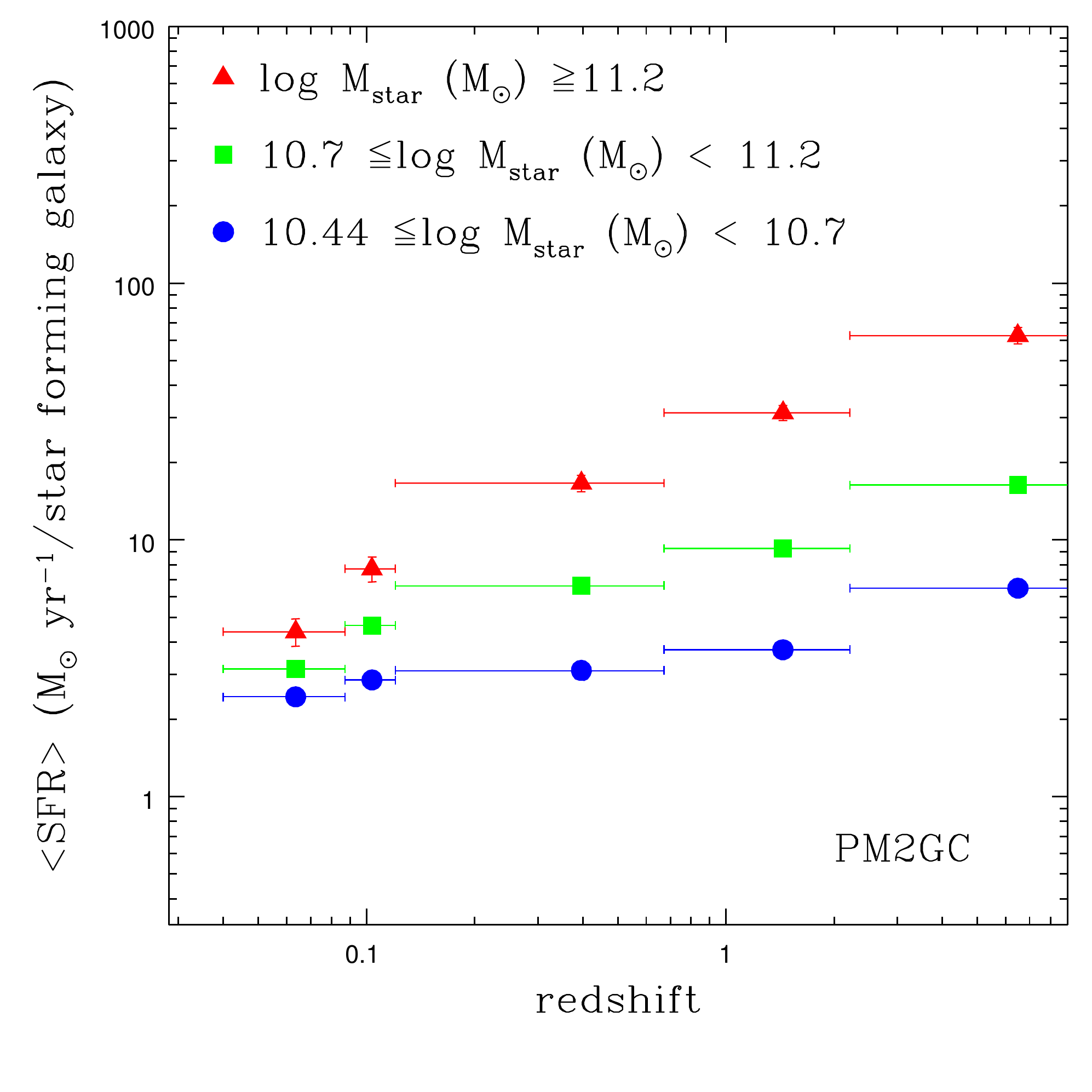}
\hspace{2mm}
\includegraphics[scale=0.425]{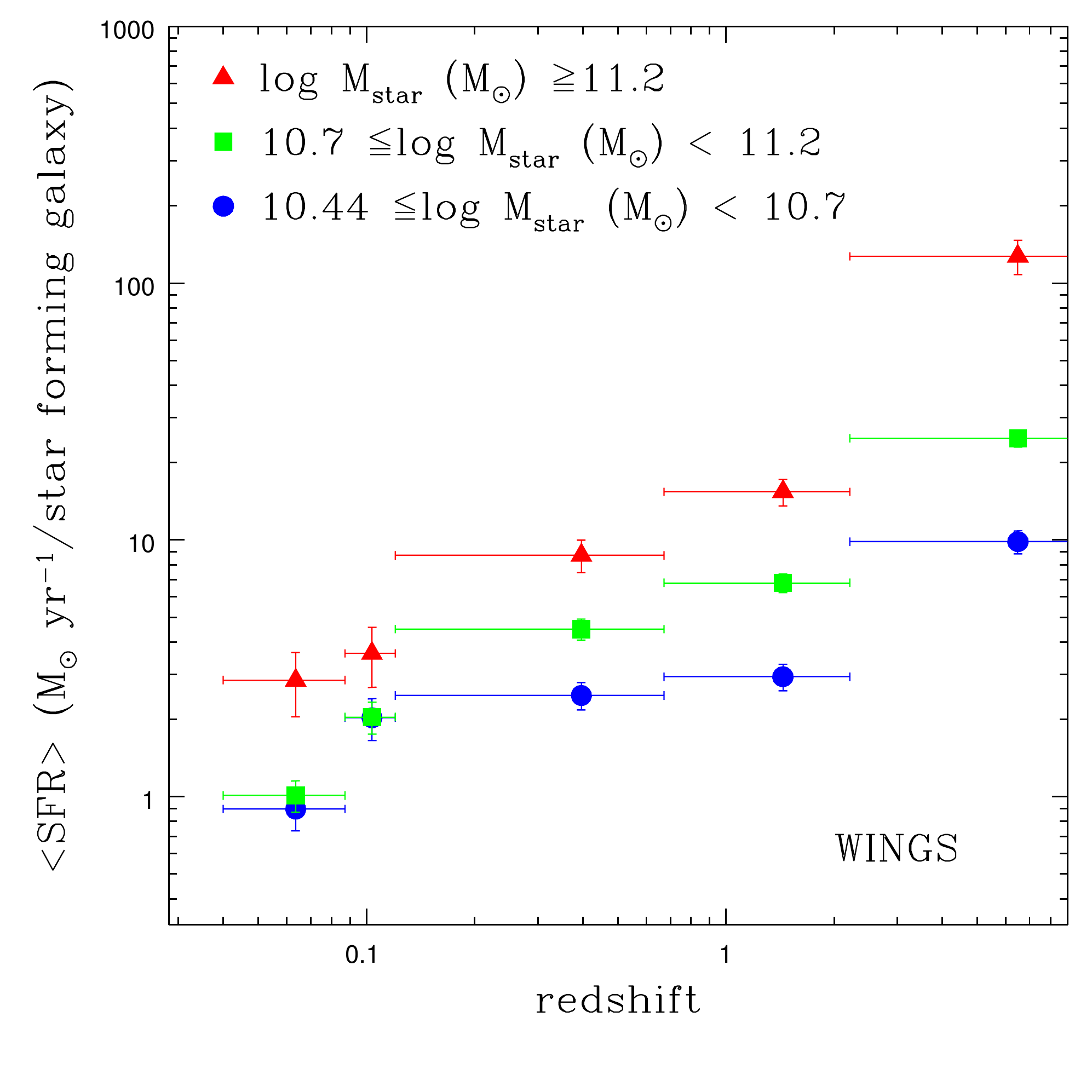}
\caption{The PM2GC (left) and WINGS (right) mean SFR of today's star-forming galaxies. Galaxies are considered as star-forming if they have a sSFR higher than  $10^{-12}$ in the last 20 Myrs. The selected galaxies are divided into three mass bins, plotted with different colours and shapes as shown in the legend. Horizontal bars refer to the time interval over which the mean SFR is computed, while vertical ones are associated with errors in the SFR determination.}
\label{SFingmass}
\end{figure*}

Figure~\ref{SFingmass} shows the mean SFH per star-forming galaxy in different galaxy mass bins, obtained dividing the sum of all SFRs by the number of galaxies.
The global decline in the cosmic star formation is not only due to an increasing fraction of galaxies becoming quenched at lower redshifts, but also to the decrease with time of 
the average SFR of today's star-forming galaxies.

The trend depends on galaxy mass, as shown in Fig. \ref{SFingmass}: it is steeper in high-mass galaxies than in low-mass ones, both in the field and in clusters. 
In clusters, the SFR drop between the oldest and the second oldest time intervals is much more pronounced than in the field for all galaxy masses, in agreement with the fact that star formation in cluster galaxies occurred very early on.

\begin{figure}
\includegraphics[scale=0.45]{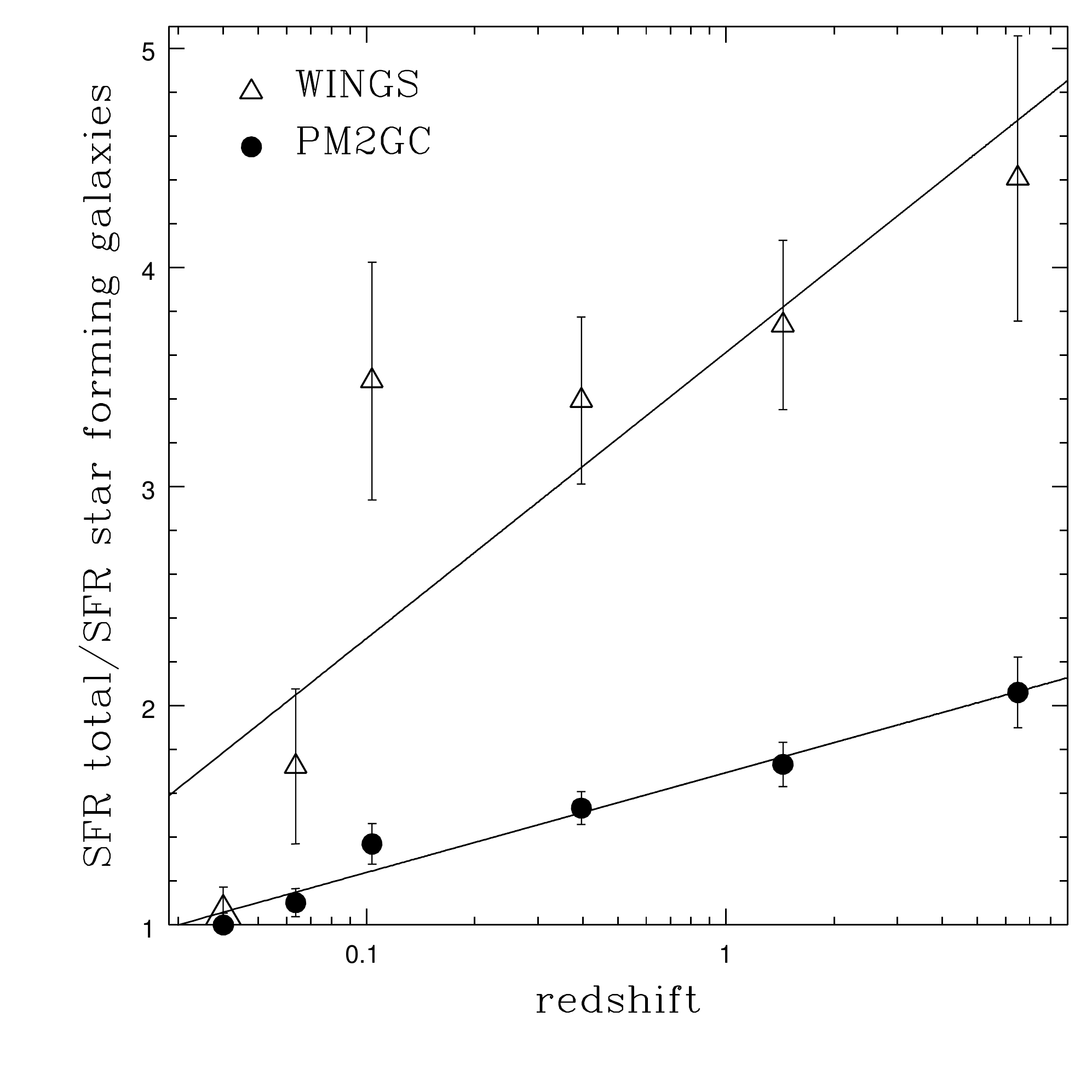}
\caption{Ratio between the SFR from the complete sample and the SFR of currently star-forming galaxies for PM2GC (full circles) and WINGS (empty triangles). The solid lines are the linear interpolation computed using an ordinary least square method and whose equations are given in eqn.~\ref{interp_eq}. Error bars have been computed from the errors relative to SFRs in both the complete and star-forming samples using error propagation.}
\label{interpolazione}
\end{figure}

Figure \ref{interpolazione} shows 
the redshift dependence of the ratio between the total SFR of all galaxies at any given redshift and the total SFR at the same redshift of galaxies that are still forming stars today, for PM2GC (full circles) and WINGS (empty triangles) separately. The fractional contribution to the total SFR at any redshift of galaxies that are now quenched is equal to (1 - 1/y), with y being the Y axis-value in Fig.~\ref{interpolazione}. There is one extra redshift bin in this figure, because the first time interval of 600 Myr was splitted into 20 Myr and 580 Myr, to isolate the present-day value according to our definition of star forming galaxies. The first point plotted in the figure represents the ratio between the SFR of today's star forming galaxies and the current measured SFR, and by definition it has a value equal to one.

The resulting values have been interpolated using a least squares method and the resulting interpolation lines are:

\begin{eqnarray}
{\rm PM2GC}: {\rm y = (0.46 \pm 0.04) \times \log(z) + (1.7 \pm 0.04)} \\
{\rm rms = 0.061} \nonumber \\
{\rm WINGS}: {\rm y  = (1.31 \pm 0.39) \times \log(z) + (3.61 \pm 0.36)}\\
{\rm rms = 0.61} \nonumber
\label{interp_eq}
\end{eqnarray}

where y is the Y axis-value in Fig.~\ref{interpolazione}.

 A good correlation for PM2GC galaxies is found, with the interpolation line reproducing well within the error bars the temporal behaviour of the ratio. For WINGS not all the points follow a linear correlation within the error bars, even if the general trend is  decreasing as in the field.

This figure illustrates that the star formation process at high z was mainly due to galaxies which today are not active anymore. In fact, $\sim 50\%$ of the SFR in the field and $\sim 75\%$ in clusters at $z>2$ originated in galaxies that are not forming stars today. At $z \sim 1.5$, these factors are $42\%$ in the field and $73\%$ in clusters.

Moreover, 
in clusters the interpolation line is almost three times steeper than in the field, meaning that the contribution of quenching to the SFRD(z) decline is much more significant in clusters than in the field.

\subsection{The contribution to the SFRD(z) of galaxies of different morphologies and masses}
 \label{subsec_morphomass_plots}

In this section we focus on the comparison between clusters and field taking into consideration the contribution to the SFRD(z) of galaxies of different mass and morphology. Recall that stellar masses and morphologies refer to galaxies as they appear at low redshift, when we observe them. Their morphological type at higher z, at the moment they possessed the SFRs we infer, might have been different, due to the well known morphological evolution taking place both in clusters and in the field (e.g. Dressler et al. 1997; Oesch et al. 2010; Vulcani et al. 2011).

\begin{figure*}
\includegraphics[scale=0.425]{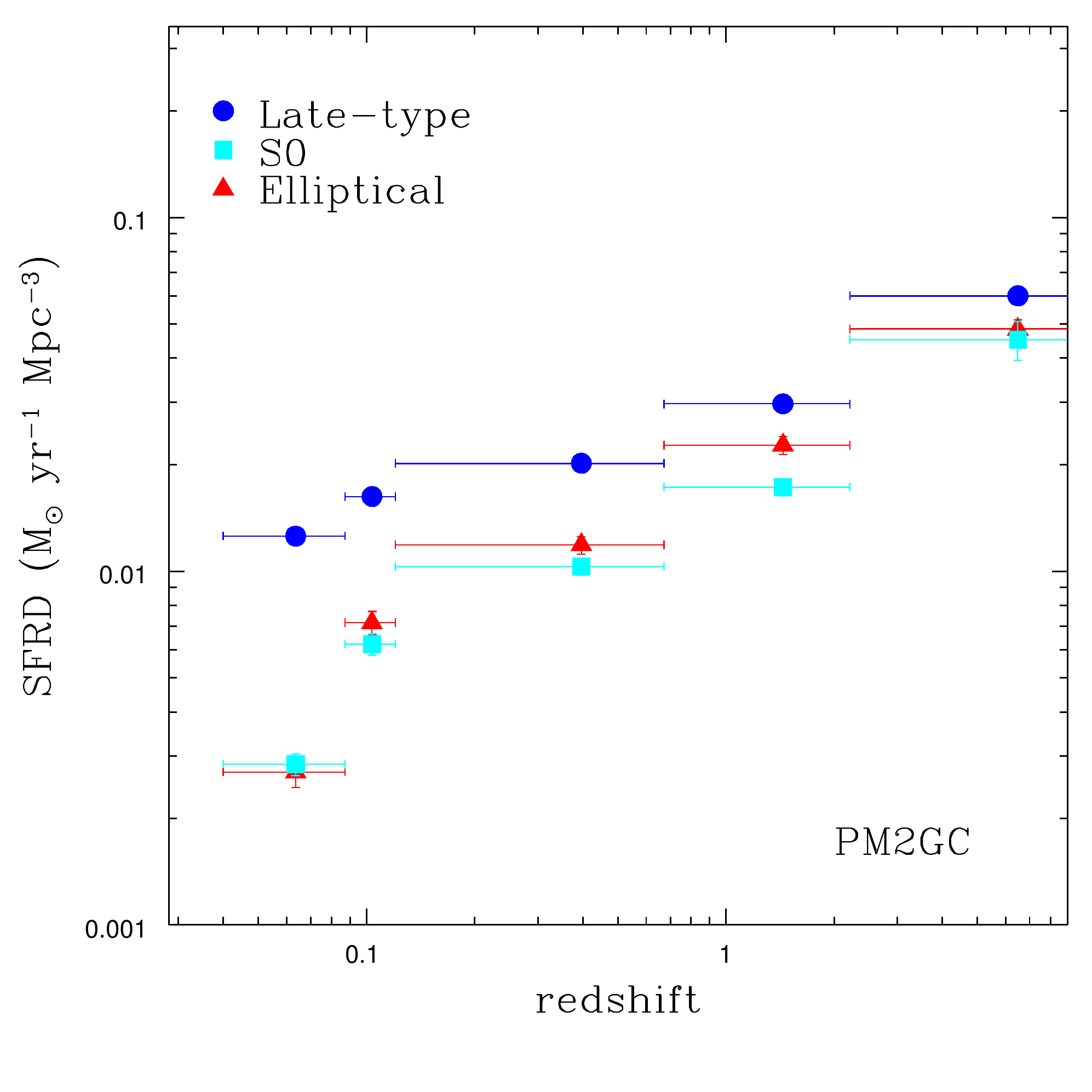}
\hspace{2mm}
\includegraphics[scale=0.425]{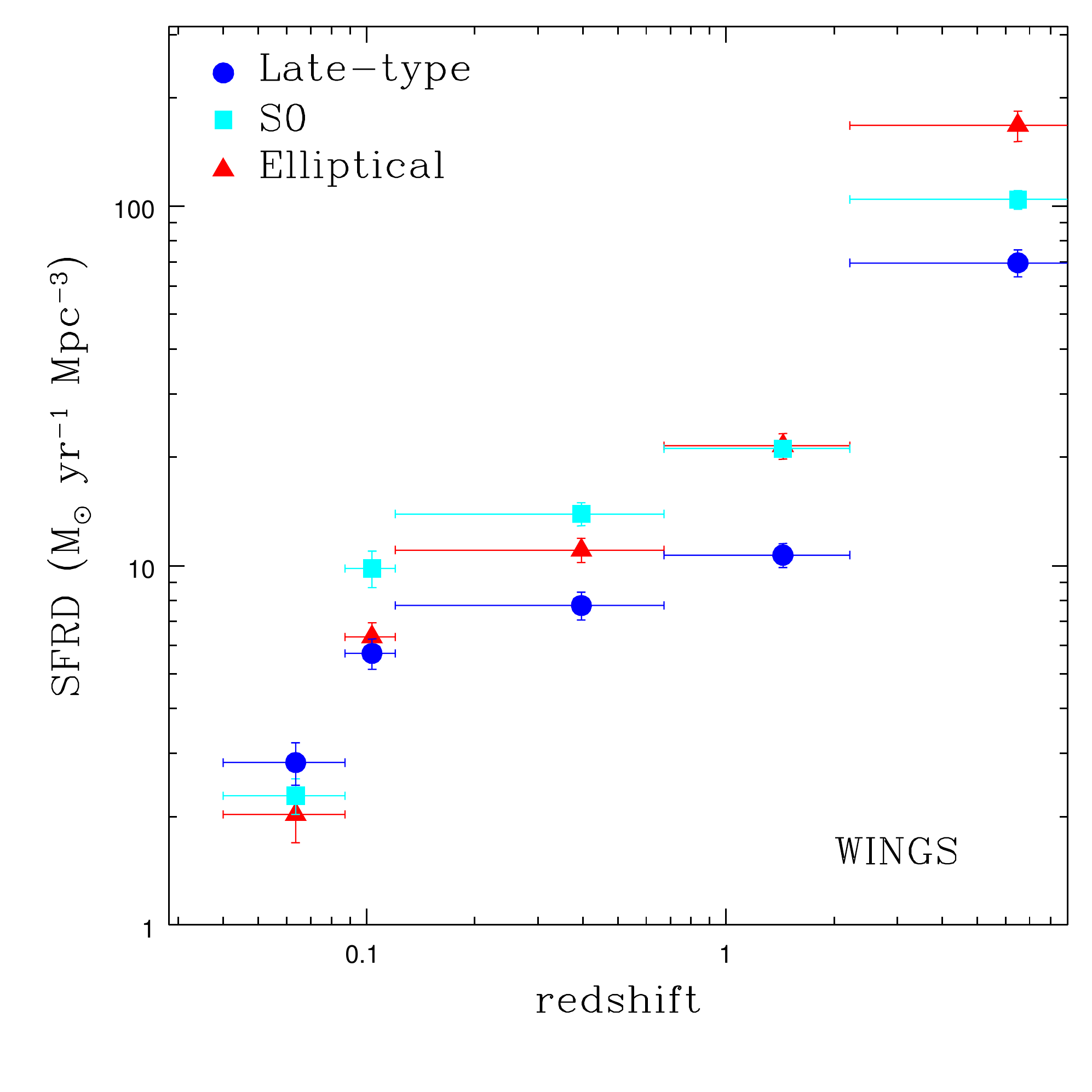}
\caption{The SFH of PM2GC (in the left panel) and WINGS (in the right panel) whose galaxies have been divided according to their morphological type. Red triangles stand for ellipticals, cyan squares for lenticulars (S0) and blue circles for late-types. All SFRDs refer to the same time intervals, here represented with horizontal bars. Vertical bars are associated to indetermination in SFRD values.}
\label{PM2wings_morpho}
\end{figure*}

Figure \ref{PM2wings_morpho} shows that the contribution to the SFRD(z) depends on the morphological type and, considering a given type, 
on the environment.\footnote{We note that the results for early-type galaxies (ellipticals and S0s) should be considered as upper limits, since the presence of AGN could produce an overestimation of the SFR,  therefore of the SFRD. Nonetheless, in our sample the total contribution from AGN is negligible, as discussed in sec.
\ref{subsec_fitting_outputs}. 
AGNs classified as early-type in the PM2GC sample are 28 and their contribution to the z=0 early-type SFRD is $\sim 5.4\%$. The number of early type AGNs contaminating WINGS galaxies is only 3: in the present epoch they contribute to the early-type SFRD only for $\sim 1.36\%$.} 

The main contribution to the SFRD in the field sample (left panel in the figure) is given by today's late-type galaxies (marked with blue circles), which dominate at all redshifts. Compared to the total values estimated for the PM2GC, the SFRD of late-types is $\sim 70\%$ of the total at the present epoch and $\sim 40\%$ at the highest z.
The relative contribution of different morphological types to the total star formation varies with time: (today's) early-type galaxies, which are composed mainly of old and red stars, gave a larger contribution to the SFRD at earlier epochs, while today they contribute only for 30 $\%$ of the total SFRD.
S0s and ellipticals have quite similar values at every epoch, with ellipticals slightly dominating at all redshifts except the lowest bin.



In principle, the analysis just performed depends on both the stellar history of each type and the morphological distribution of galaxies within each environment, i.e. the number of galaxies populating each type. In our field sample 59\% of all galaxies are late-types, 21\% are S0s and 19\% are ellipticals. At z = 0 on average the star formation activity in a late-type galaxy is 1.5 times higher than in an elliptical and 1.6 times than in an S0, which is expected given that early-type galaxies today are on average more passive than late-types.

In contrast with the field, early-type galaxies dominate the total SFRD in clusters at all epochs, except in the lowest redshift bin. The difference in the fractional contribution of late- and early-types, however, is much smaller than in the field: in clusters, 40 $\%$ of the total today SFRD is due to late-types, only slightly higher than the 32 $\%$ and 28 $\%$ of S0s and ellipticals, respectively.
This picture reverses going back in time: between 0.1 $\lesssim$ z $\lesssim$ 1, today's lenticular galaxies produce the majority of stars, and finally at the highest z ellipticals dominate.
The scenario just described is influenced by the significantly different distribution of morphologies in cluster galaxies compared with that in the field: in clusters 28\% of all galaxies are ellipticals, 44\% are S0s and 27\% are late-type. 

Overall, the trends in clusters and in the field are clearly very different as far as the relative roles of each type are concerned.

\begin{figure*}
\includegraphics[scale=0.425]{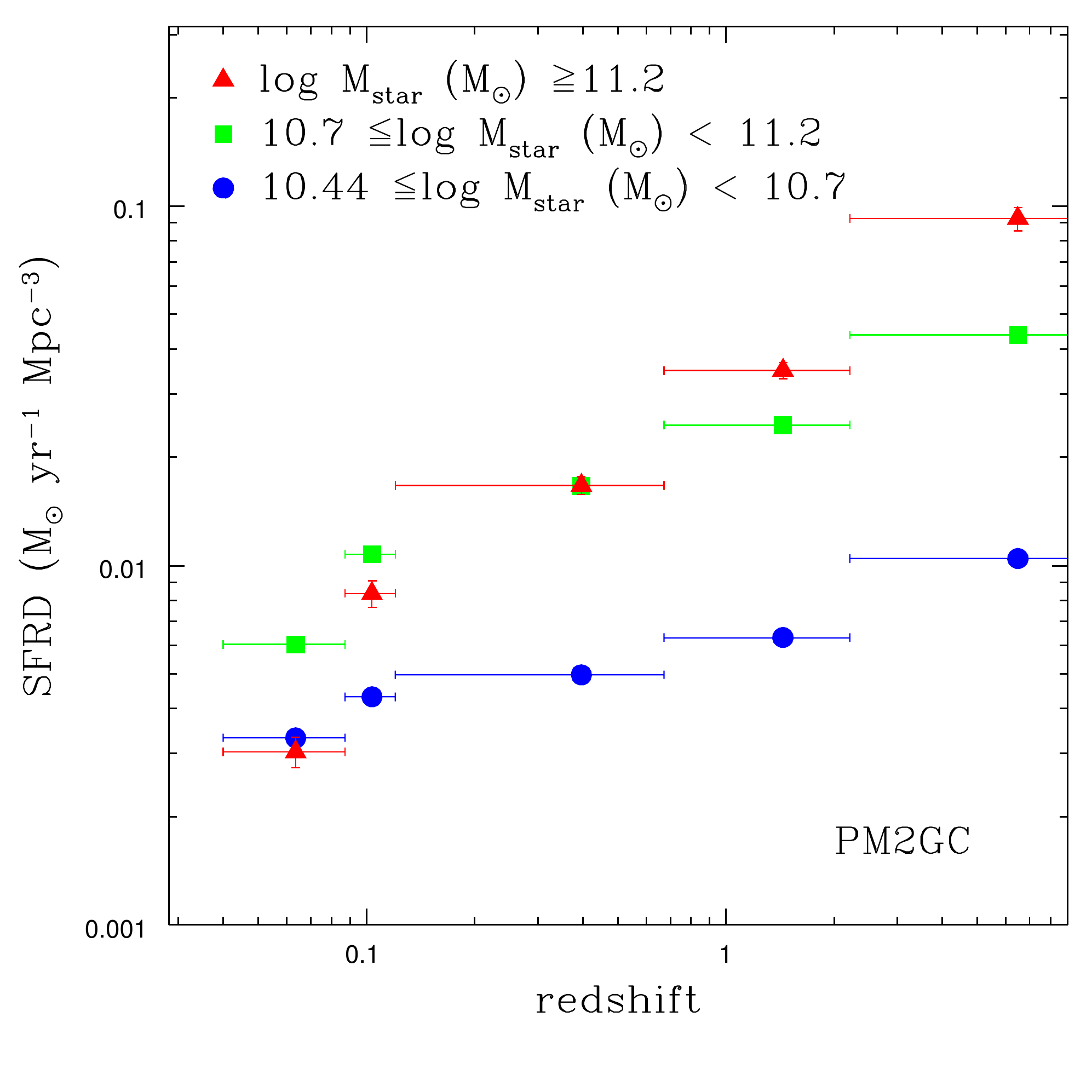}
\hspace{2mm}
\includegraphics[scale=0.425]{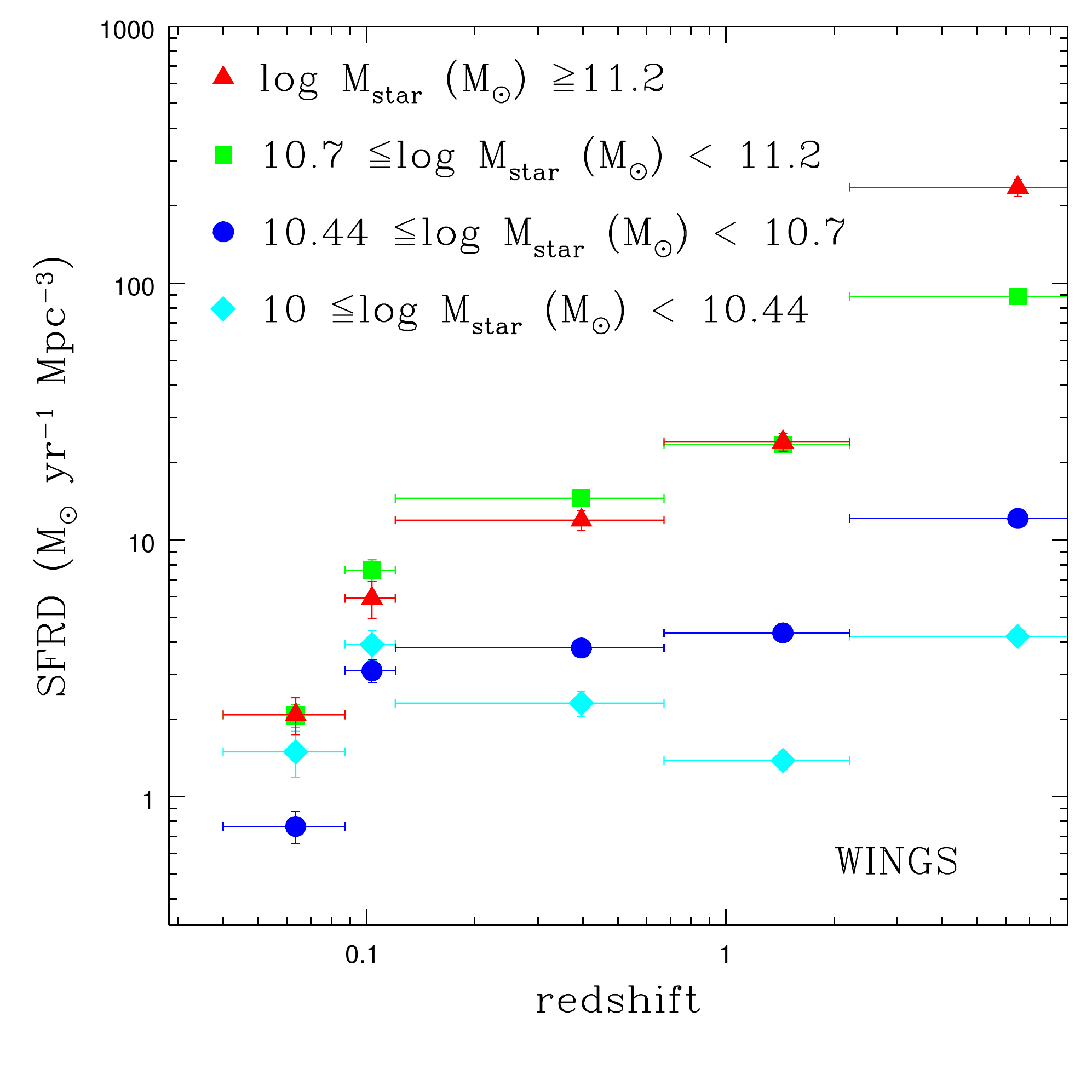}
\caption{The PM2GC (on the left) and WINGS (on the right) SFH for galaxies divided in mass bins, as shown in the legend. Mass values here reported are calculated according to a Salpeter (0.1-100) IMF. All SFRDs are supposed to be constant in the same time intervals, here represented with horizontal bars. Vertical bars are associated to indetermination in SFRD values.}
\label{pm2wings_mass}
\end{figure*}

We now divide galaxies into mass bins, according to the completeness limits of the two surveys: M $\geqslant 10^{10.44} M_{\odot}$ for PM2GC and M $\geqslant 10^{10} M_{\odot}$ for WINGS. 
In figure \ref{pm2wings_mass} the field and the cluster SFRDs are divided into respectively three and four mass bins. 
Qualitatively, the global SFRD in both environments is dominated by galaxies with $M >10^{10.7} M_{\odot}$.
Going into more details, in the field, galaxies with masses M $\geqslant 10^{11.2} M_{\odot}$ give the main contribution to the total SFRD for z $\gtrsim$ 0.35, while in the two lowest redshift intervals galaxies with masses $10^{10.7} M_{\odot} \leqslant M < 10^{11.2} M_{\odot}$ prevail. Low-mass galaxies (blue circles) have lower SFRD than the intermediate mass galaxies, but still higher than the most massive galaxies for z $\lesssim$ 0.1.
The trends plotted in the left panel of figure \ref{pm2wings_mass} are influenced again by both the size of the subsamples and the average SFRs of galaxies of different masses.

To compare the average SFR of a typical galaxy of a certain mass it is useful to divide the SFR by the number of galaxies populating the considered mass bin. Today, on average, galaxies in different mass bins (including passive ones) form roughly the same amount of stars, and intermediate-mass galaxies dominate the global SFRD in the field just because they are more numerous. On the contrary, at higher z, the hierarchy established in the total SFRD is very pronounced: the SFR per unit galaxy of M $\geqslant 10^{11.2} M_{\odot}$ galaxies on average is $\backsim$ 5 times higher than that of intermediate-mass galaxies and about 13 times the one of the low mass systems.

The right panel of figure \ref{pm2wings_mass} refers to WINGS cluster galaxies. 
Galaxies with the lowest mass always give the smallest contribution to the total SFRD at any redshift, while the intermediate and high-mass galaxies have similar SFRD values until z $\backsim$ 2, and at higher z the most massive ones prevail.
Analysing again the mean SFRD per galaxy within a certain range in mass it turns out that today the SFRD of the average low-mass galaxy becomes roughly equal to that of intermediate-mass ones, while high-mass galaxies have values higher of a factor $\backsim$ 2.3. Going to higher redshifts, the ratio of the average SFRD of high-mass and intermediate-mass (low-mass) galaxies becomes greater, reaching a value of $\backsim$ 6 (16) at the highest redshift.

An important phenomenon strictly connected with masses is downsizing: galaxies with higher masses are characterized by shorter and earlier star formation on average, while lower mass galaxies have longer star formation timescales. The variation of the SFH of galaxies according to their stellar mass is evident comparing the slopes of the hypothetical curve connecting points of the same color and shape (red, green and blue) in Fig. \ref{pm2wings_mass}. 
In particular, as an estimate of the process, we can calculate the ratio between the value of SFR in the first and in the last redshift intervals and analyse its variation as a function of mass.
This ratio is $\backsim$ 3 for low-mass galaxies, $\backsim$ 6 for intermediate mass ones and $\backsim$ 26 for high mass galaxies in the field. The same ratios in WINGS galaxies are the following: $\backsim$ 13 for low masses, $\backsim$ 36 for intermediate ones and $\backsim$ 94 for the highest masses. The numbers listed above demonstrate that the downsizing phenomenon acts in the field as well as in clusters, but in the latter it is stronger. Even galaxies of the same mass are characterised by different timescales and SFHs depending on their environment. The average decline of the star formation process in galaxies of a given mass is less steep in the field than in clusters. Cluster galaxies form the bulk of their stars at earlier epochs with average high-z SFRD values per galaxy systematically higher than those of the field at any given galaxy stellar mass today, as we will see in more details in the next section. 

\subsection{Masses, morphologies and environment: a global picture}

\begin{figure*}
\includegraphics[scale=0.425]{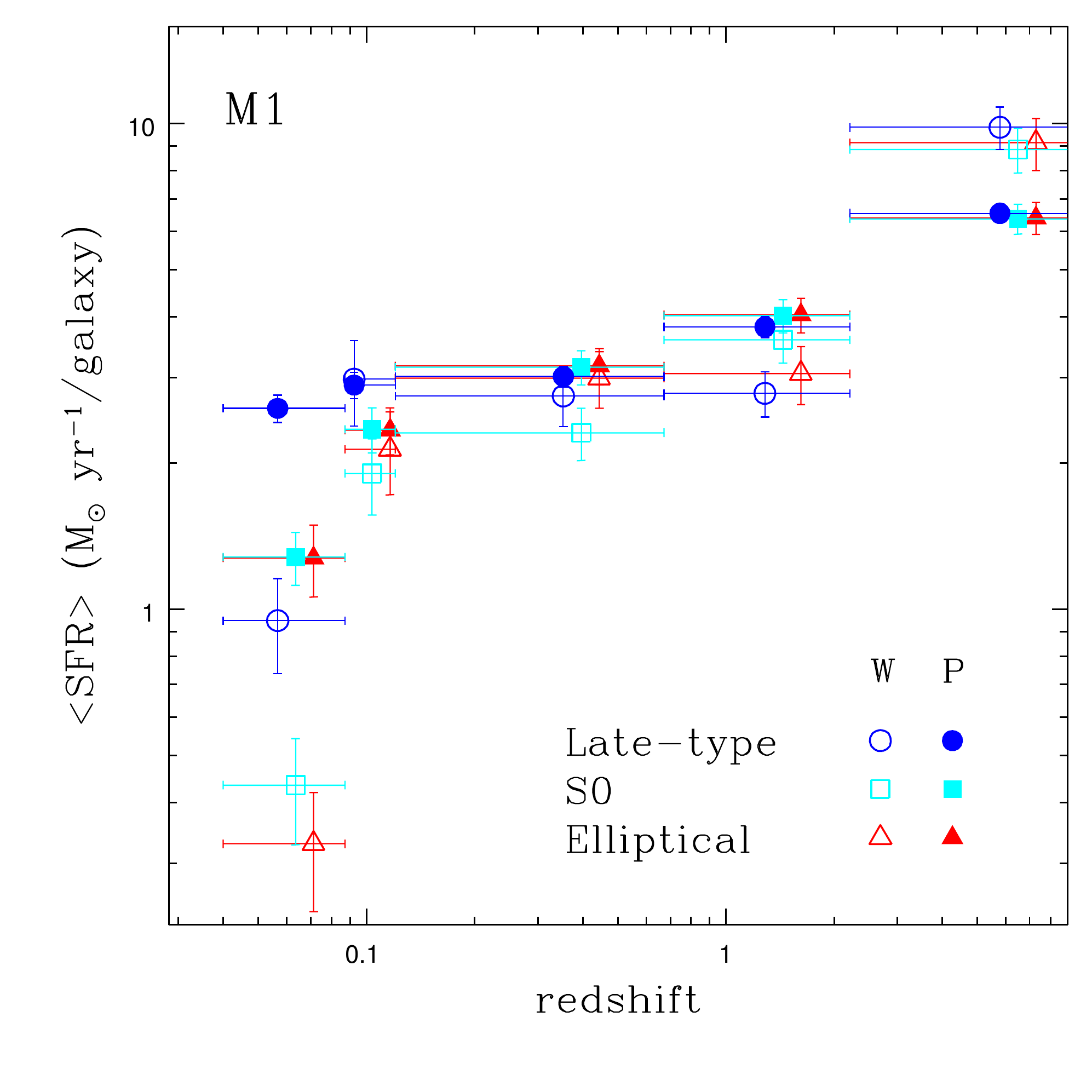}
\hspace{1mm}
\includegraphics[scale=0.425]{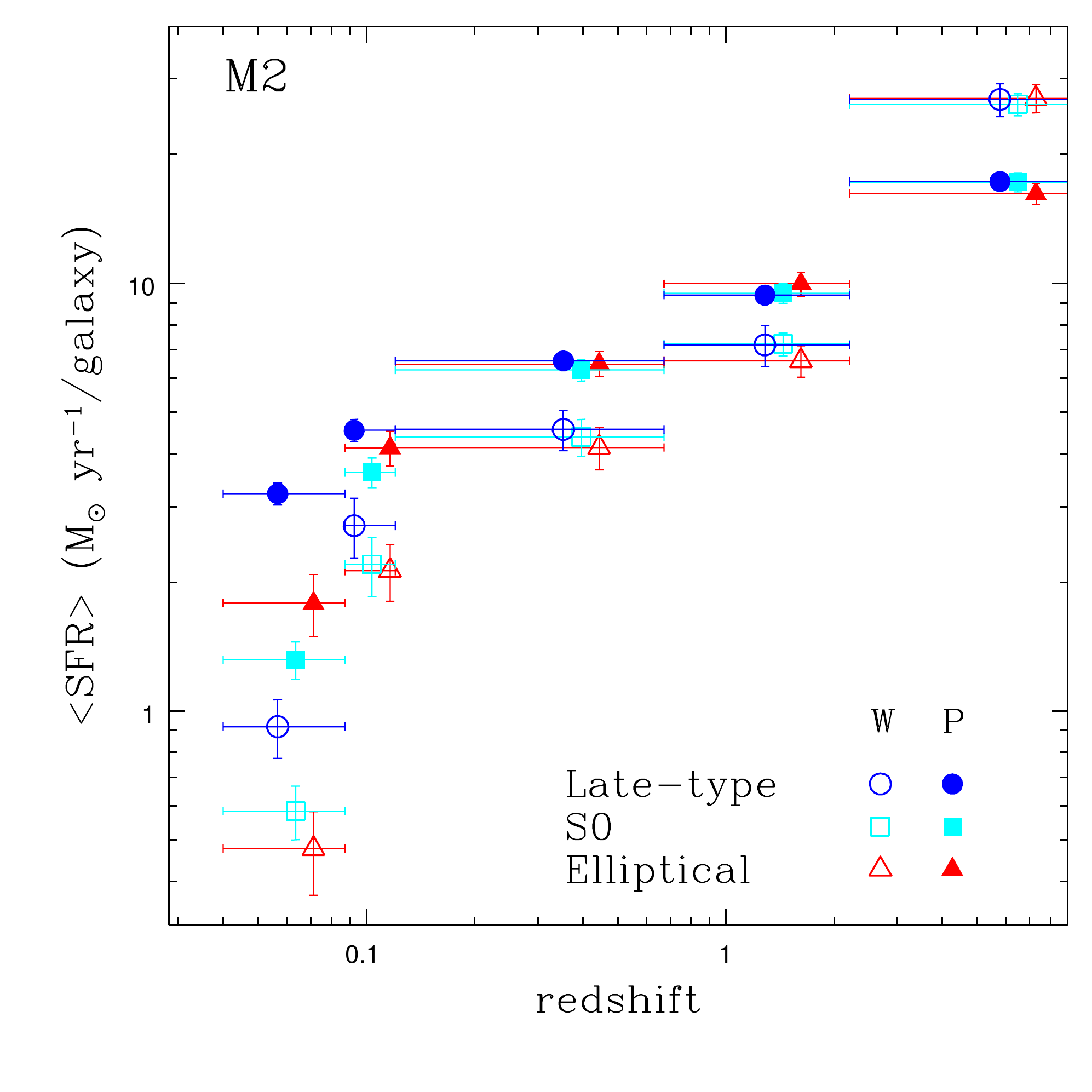}
\hspace{1mm}
\includegraphics[scale=0.425]{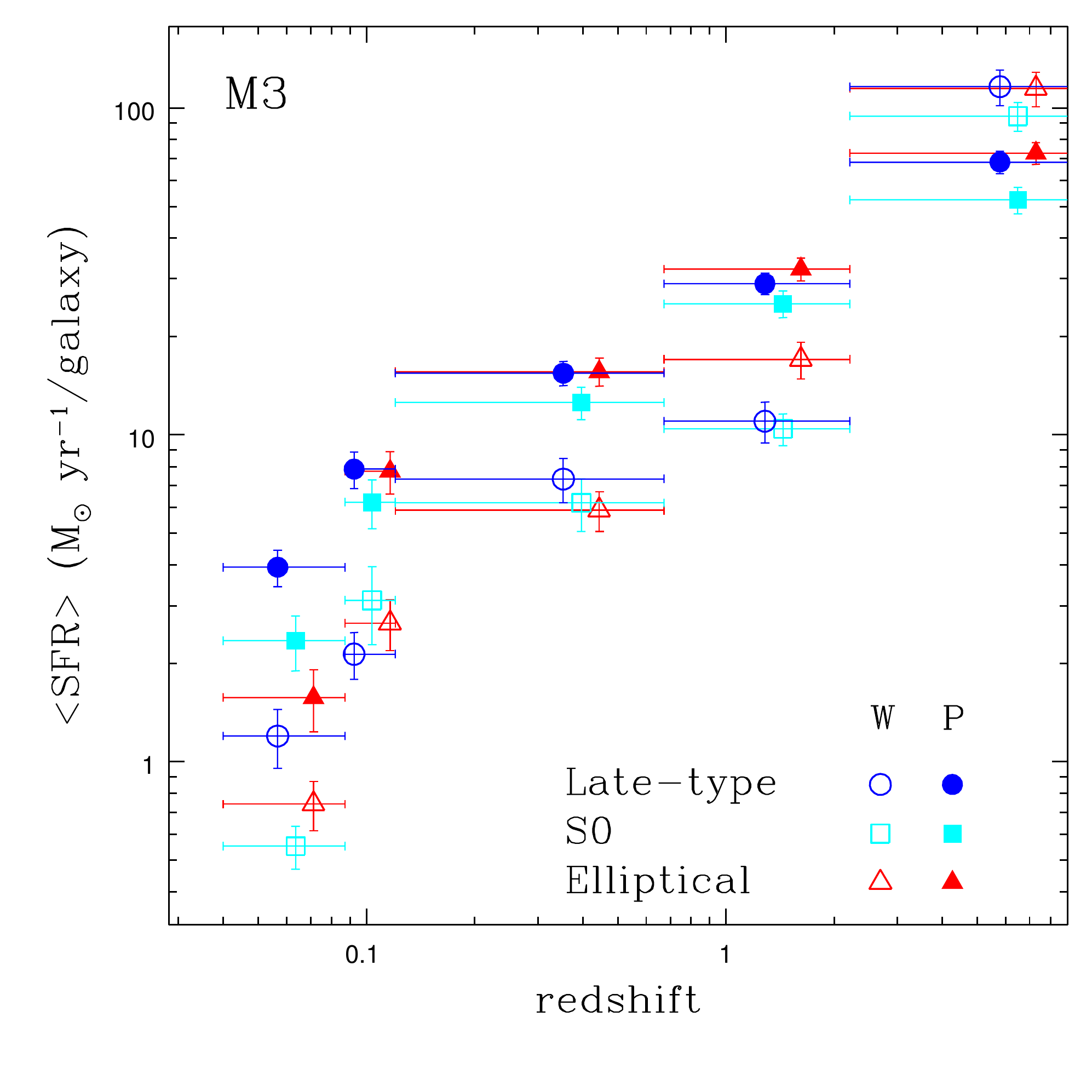}
\caption{The SFH of galaxies with different morphological type and mass: in the first plot galaxies have $10^{10.44} M_{\odot}\leq M1<10^{10.7}M_{\odot}$, in the second galaxies have $10^{10.7} M_{\odot}\leq M2 <10^{11.2}M_{\odot}$ and in the third galaxies have masses $M3\geq 10^{11.2}M_{\odot}$. Data reported with full circles, squares and triangles refer to the field sample PM2GC and empty ones refer to the cluster sample WINGS, with different colours meaning different morphological types, as shown in the legend. The average SFRs are assumed to be constant in the same temporal extension, here represented with horizontal bars. Vertical bars are associated to indetermination in SFR values. Symbols are horizontally shifted by small arbitrary amounts within their redshift bin in order to avoid superpositions.}
\label{massM123}
\end{figure*}

The last sequence of plots in figure \ref{massM123} aims to answer the following questions: on average, do galaxies of different morphological types but same masses have different histories? Do galaxies of the same mass and morphological type have different histories depending on the environment?

We divide galaxies according to their morphological type, mass bin and environment.
To avoid any possible residual mass dependence in each mass bin, we first verified whether galaxies in each given mass bin and given morphological type had the same mass distribution in both environments,
performing a Kolmogorov Smirnov test (KS).
The KS test found significantly different mass distributions only in three cases
(elliptical galaxies in the lowest mass bin and lenticular galaxies both in the lowest and in the highest mass bins, plots not shown).
For these, we constructed ad hoc samples of randomly selected WINGS galaxies that matched the PM2GC mass distribution. Moreover, for this plot, we limit the M3 bin to $< 7 \times \, 10^{11} M_{\odot}$, to have a similar upper mass limit for spirals, ellipticals and S0s in each environment.

Fig. \ref{massM123} presents the SFH of galaxies in
each mass bin, matched in mass when necessary, for different environments and morphologies. The total SFR is divided by the number of galaxies of each subsample, in order to derive the mean history of a galaxy of a given type, mass and environment. 
Environments are plotted in figures with different symbols (full circles, squares and triangles for field galaxies and empty circles, squares and triangles for cluster galaxies) and colours follow the same legend of the morphological analysis in figure \ref{PM2wings_morpho}.

Figure~\ref{massM123} highlights that, perhaps surprisingly, in a given environment, galaxies of the same mass but different morphologies share the same history of star formation, except for the lowest redshift bin. In fact, the average SFR of late-type, S0 and elliptical galaxies of a given mass is similar within the errors at all redshifts, except at $z<0.1$ when late-type galaxies have a systematically higher value than early-type galaxies.
There is, instead, a different SFH in clusters and field for galaxies of a given mass and morphological type: all types in clusters have a higher SFR at $z>2$ and a lower SFR at lower redshifts, than field galaxies.



Comparing now the different mass bins, the downsizing in star formation is again visible (the slope of the SFH gets steeper at higher masses). The range of SFR at the lowest redshift is similar for galaxies of all masses, while it varies greatly at the highest z (see Sect. \ref{subsec_morphomass_plots}).

From the analysis of these plots, it is possible to establish a general hierarchy in the properties of galaxies which mostly influence the star formation process.

Galaxy mass is clearly an important factor in determining the SFR slope with time, in all environments. However, it is not the only factor, as cluster galaxies of a given mass have a steeper decline of SF with time. For each mass, the highest average SFR at $z>2$ is found for cluster galaxies, and does not depend on morphology.

Morphology, in contrast, has little influence on the SFH. The only morphological dependence is at the present epoch when, on average, a late-type galaxy forms more stars than an early-type galaxy of the same mass.

These results seem to suggest that the stellar history of a galaxy depends mainly on its mass \emph{and} environment, and is almost independent of its present-day morphology. 

Finally, computing from Fig.~\ref{massM123} the ratio of the average SFR in the highest and lowest redshift bins for galaxies of the same mass and morphology and comparing it for different environments, we obtain cluster-to-field ratios typically ranging from 4 to 7, with an average of 5. Thus,
the much steeper SFRD decline in clusters compared to the field (ratio=40/7=5.7) discussed in \ref{subsec_environ_plots} can be explained by the steeper history of cluster galaxies compared to the field, at fixed galaxy mass and morphology. We conclude that the different slope in the SFRD(z) of clusters and field is not driven by variations of the galaxy mass or morphological distributions with environment, but by the fact that galaxy stellar histories vary with galaxy location at each given mass and morphology.

\section{Summary}

Having derived the SFH of galaxies in clusters (WINGS) and the field (PM2GC), we have investigated the SFRD evolution with redshift as a function of environment, the histories of galaxies that are still forming stars at the time they are observed, and the role of galaxy masses, morphologies and environment in driving the differences of the SFRD(z) with environment.
We have found that:
\begin{enumerate}
\item The PM2GC cumulative SFRD agrees quite well with the SFRD observed at different redshifts (figure \ref{Pm2gc_MD14}). The only discrepancy is seen at the highest z (z $>$ 2) where the PM2GC SFRD is a factor $\sim 1.7$ lower than the integral of the MD14 best-fit function over the same redshift interval.
\item The SFRD changes with environment (figure \ref{pm2wings_subsets}) and in particular two effects contribute simultaneously to the cluster-field differences in the SFRD: the different density of the two environments, which changes the normalization of the SFRDs at all epochs, and the intrinsic differences of the histories within the environments, which change the slope of the SFRD. The cluster SFRD decline is much steeper than in the field, and there is a progressive steepening going from single to binaries to groups and clusters, as well as going from lower mass to higher mass haloes.
\item The decline of the SFRD(z) is due to two factors: the decline of the SFH of star forming galaxies and the quenching rate of galaxies as a function of redshift (figure \ref{SFingmass}). We have quantified the relative importance of the two processes (figure \ref{interpolazione}): the star formation process at high z was mainly due to galaxies which today are not active anymore, and this is true in particular for clusters. More than 50 \% of the SFR in the field and more than 75 \% in clusters at z $>$ 2 originated in galaxies that are not currently forming stars. At z $\sim$ 1 these factors are 42 \% in the field and 73 \% in clusters. 
\item 
Galaxies of different morphological types but same mass and environment have on average remarkably similar SFRs at all epochs except at the lowest redshift,
suggesting that the current morphological type is linked with the current morphology but is largely non-influent for the past SFH. 
\item The average SFH of a galaxy depends on galaxy stellar mass and, at fixed mass, on galaxy environment. The different slope of the decline in the SFRD(z) in clusters and field is due to the fact that galaxies of given mass and morphology form their stars sooner in clusters than in the field. 
\end{enumerate}
These results point to an accelerated formation in high mass haloes compared to low mass ones even for galaxies that will end up having the same galaxy mass today.



\section*{Acknowledgments}
We acknowledge the anonymous referee for her/his careful report, important suggestions and comments which helped us improving our work.
We thank Joe Liske, Simon Driver and the whole MGC team for making easily accessible a great data set. We are grateful to the rest of the WINGS team for help and useful discussions.
VG and BMP acknowledge financial support from the Istituto Nazionale di Astrofisica through a PhD Cycle 30th grant.
BV was supported by the World Premier International Research Center Initiative (WPI), MEXT, Japan and by the Kakenhi Grant-in-Aid for Young Scientists (B)(26870140) from the Japan Society for the Promotion of Science (JSPS).

We gratefully acknowledge Prof. Giuseppe Tormen for inspiring discussions, precious suggestions and support as internal supervisor during all the work.

\bibliographystyle{alpha}

\appendix

\bsp

\label{lastpage}

\end{document}